\documentclass[a4paper,USenglish,cleveref, autoref, thm-restate,runningheads]{llncs}
\newcommand{\longversion}[1]{#1}
\newcommand{\shortversion}[1]{}
\longversion{\pdfoutput=1} 

\shortversion{\usepackage{lineno}\linenumbers}

\usepackage[dvipsnames]{xcolor}
\usepackage{todonotes}
\usepackage{comment}

\newcommand{\strhf}[1]{\shortversion{}\longversion{\todo{\color{blue}#1}}}

\longversion{\usepackage[appendix=inline]{apxproof}}
\shortversion{\usepackage{apxproof}}%

\newtheoremrep{theorem}{Theorem}[section] 
\newtheoremrep{proposition}[theorem]{Proposition}
\newtheoremrep{lemma}[theorem]{Lemma}
\newtheoremrep{claim}[theorem]{Claim}
\newtheoremrep{observation}[theorem]{Observation}
\newtheoremrep{conjecture}[theorem]{Conjecture}
\newtheoremrep{corollary}[theorem]{Corollary}
\theoremstyle{definition}
\newtheoremrep{definition}[theorem]{Definition}
\theoremstyle{remark}
\newtheoremrep{example}[theorem]{Example}
\newtheoremrep{remark}[theorem]{Remark}

\usepackage[utf8]{inputenc}          
\usepackage[T1]{fontenc}             
\usepackage{amsfonts}                
\usepackage{amssymb}                 
\usepackage{amsmath}                 

\usepackage{mathtools}
\usepackage{setspace}
\usepackage{graphicx}
\usepackage{hyperref}
\usepackage{xcolor}
\usepackage{xspace}
\usepackage{subcaption}

\usepackage{algorithm}
\usepackage[noend]{algpseudocode}

\newcommand{\DSReconf}{\longversion{\textsc{Dominating Set Re\-con\-figur\-ation}\xspace}\shortversion{\textsc{DS-Reconf}\xspace}}
\newcommand{\AllReconf}{\longversion{\textsc{Alliance \linebreak[3] Re\-con\-figur\-ation}\xspace}\shortversion{\textsc{All-Reconf}\xspace}}
\newcommand{\defallianceReconf}{\longversion{\textsc{Defensive Alliance Re\-con\-figur\-ation}\xspace}\shortversion{\textsc{DA-Reconf}\xspace}}
\newcommand{\offallianceReconf}{\longversion{\textsc{Offensive Alliance Re\-con\-figur\-ation}\xspace}\shortversion{\textsc{OA-Reconf}\xspace}}

\longversion{\newcommand{\offall}{offensive alliance\xspace}
\newcommand{\defall}{defensive alliance\xspace}
\newcommand{\powall}{powerful alliance\xspace}
\newcommand{\offalls}{offensive alliances\xspace}
\newcommand{\defalls}{defensive alliances\xspace}
\newcommand{\powalls}{powerful alliances\xspace}
\newcommand{\defalltslideseq}{\defall token sliding sequence\xspace}
\newcommand{\defalltjumpseq}{\defall token jumping sequence\xspace}

\newcommand{\alltjumpseq}{alliance token jumping sequence\xspace}
\newcommand{\alltarseq}{alliance token addition removal sequence\xspace}

\newcommand{\defalltarseq}{\defall token addition removal sequence\xspace}
\newcommand{\offalltslideseq}{\offall token sliding sequence\xspace}
\newcommand{\offalltjumpseq}{\offall token jumping sequence\xspace}
\newcommand{\offalltarseq}{\offall token addition removal sequence\xspace}

\newcommand{\dstjumpseq}{dominating set token jumping sequence\xspace}

}

\shortversion{\newcommand{\defall}{DA\xspace}
\newcommand{\offall}{OA\xspace}
\newcommand{\powall}{PA\xspace}
\newcommand{\defalls}{DAs\xspace}
\newcommand{\offalls}{OAs\xspace}
\newcommand{\powalls}{PAs\xspace}
\newcommand{\defalltslideseq}{\defall TS sequence\xspace}
\newcommand{\defalltjumpseq}{\defall TJ sequence\xspace}

\newcommand{\alltjumpseq}{alliance TJ sequence\xspace}
\newcommand{\alltarseq}{alliance TAR sequence\xspace}

\newcommand{\defalltarseq}{\defall TAR sequence\xspace}
\newcommand{\offalltslideseq}{\offall TS sequence\xspace}
\newcommand{\offalltjumpseq}{\offall TJ sequence\xspace}
\newcommand{\offalltarseq}{\offall TAR sequence\xspace}

\newcommand{\dstjumpseq}{dominating set TJ sequence\xspace}

}
\newcommand{\rmi}{\longversion{reconfiguration monotone increasing}\shortversion{rmi}\xspace}
\newcommand{\rmd}{\longversion{reconfiguration monotone decreasing}\shortversion{rmd}\xspace}

\newcommand{\NP}{\textsf{NP}\xspace}
\newcommand{\FPT}{\textsf{FPT}\xspace}
\newcommand{\XP}{\textsf{XP}}
\newcommand{\XL}{\textsf{XL}}
\newcommand{\XNL}{\textsf{XNL}}
\newcommand{\XNLP}{\textsf{XNLP}}
\newcommand{\W}[1]{\ensuremath{\textsf{W}[#1]}}

\newcommand{\pspace}{\textsf{PSPACE}\xspace}
\newcommand{\logspace}{\textsf{Log\-SPACE}}

\newcommand{\iffl}{if\longversion{ and only i}f }

\newcommand{\no}{\textsf{no}\xspace}
\newcommand{\yes}{\textsf{yes}\xspace}

\newcommand{\Oh}{\mathcal{O}}

\newcommand{\nd}{\textsf{nd}}

\newenvironment{pf}{\begin{proof}}{\hfill\qed
\end{proof}}
\newenvironment{pfclaim}{\begin{proof}}{\renewcommand{\qedsymbol}{$\Diamond$}\hfill$\Diamond$
\end{proof}}

\newcommand{\hf}[1]{\todo[color=red!27]{HF: #1}}\makeatletter
\newcommand{\crossout}[1]{%
  \begingroup
  \sbox\z@{#1}%
  \dimen\z@=\wd\z@
  \dimen\tw@=\ht\z@
  \dimen\z@=.99626\dimen\z@   
  \dimen\tw@=.99626\dimen\tw@ 
  \edef\co@wd{\strip@pt\dimen\z@}
  \edef\co@ht{\strip@pt\dimen\tw@}
  \leavevmode
  \rlap{\pdfliteral{q 1 J 0.4 w 0 0 m \co@wd\space \co@ht\space l S Q}}%
  \rlap{\pdfliteral{q 1 J 0.4 w 0 \co@ht\space m \co@wd\space 0 l S Q}}%
  #1%
  \endgroup
}
\makeatother

\begin{document}

\title{How to Reconfigure Your Alliances}
\author{Henning Fernau\inst1\orcidID{0000-0002-4444-3220}
\and Kevin Mann\inst1\orcidID{0000-0002-0880-2513} 
}
\institute{
Universit\"at Trier, Fachbereich~4 -- Abteilung Informatikwissenschaften\\  
54286 Trier, Germany.\\
\email{\{fernau,mann\}@uni-trier.de}
}
\authorrunning{H. Fernau and K. Mann}

\maketitle
\begin{abstract}
Different variations of alliances in graphs have been introduced into the graph-theoretic literature about twenty years ago. More broadly speaking, they can be interpreted as groups that collaborate to achieve a common goal, for instance, defending themselves against possible attacks from outside. In this paper, we initiate the study of reconfiguring alliances. This means that, with the understanding of having an interconnection map given by a graph, we look at two alliances of the same size~$k$ and investigate if there is a reconfiguration sequence (of length at most~$\ell$)  formed by alliances of size (at most)~$k$ that transfers one alliance into the other one. Here, we consider different (now classical) movements of tokens: sliding, jumping, addition/removal. We link the latter two regimes by introducing the concept of reconfiguration monotonicity.  Concerning classical complexity, most of these reconfiguration problems are \pspace-complete, although some are solvable in \logspace. We also consider these reconfiguration questions through the lense of parameterized algorithms and prove various \FPT-results, in particular concerning the combined parameter $k+\ell$ or neighborhood diversity together with $k$ or neighborhood diversity together with $k$.
\end{abstract}

\section{Introduction}

Abstractly speaking, the concept of \emph{reconfiguration} addresses the question how different solutions to a problem relate to each other in the sense that it is possible to `move' from one solution to another one through the space of solutions. For instance, if you do some re-installment of infrastructure, there is a working solution at present and a hopefully working solution in the future, but also all intermediate steps should be planned in a way that the infrastructure is still working for everybody. A concrete instantiation of this setting was investigated in \cite{Itoetal2011} as the \textsc{Power Supply Reconfiguration} problem. Further practically relevant examples can be found in \cite{Heu2013,Mou2015}, to cite just two references, and the current paper will add to this list of relevant problems.
Again more abstractly speaking, this type of analysis can be undertaken for any combinatorial problem. For graph problems like \textsc{Independent Set}, a reconfiguration instance would consist of a graph~$G$ and two solutions $I_s$ and $I_t$, i.e., independent sets, and the question is whether one can move from $I_s$ to $I_t$ in the solution space. In other words, the question is if there exists a \emph{reconfiguration sequence} from $I_s$ to $I_t$, formally treated in the next section.
This of course depends on the `connection structure' of the solution space. Typically, an adjacency relation between two solutions is defined based on a notion of `permitted transformation'. In this context, we imagine a solution as given by a set of tokens placed on the vertices of a graph.
For instance, \emph{token sliding} then means that two solutions $S$, $S'$ are adjacent if $S\triangle S'=\{u,v\}$, $\vert S \vert = \vert S' \vert$ and $u,v$ are adjacent in the graph, while \emph{token jumping} would not require $u,v$ to be adjacent. Similarly, one can think of \emph{token removal} or 
\emph{token addition}, to give two more examples of such `move' operations. Also, apart from the pure reconfigurability question, which is basically the question of reachability within the solution graph, one could also add a time upper bound, or just ask the combinatorial question if the solution graph is connected. 
A lot of work on various aspects of reconfiguration has been done in recent years; still, a nice introduction in the topic can be found in~\cite{Nis2018}. It should be mentioned that we assume that only one token can be at a vertex in one point of the sequence. This is not the case for each paper (for example~\cite{BonDorOuv2021}). As most variants of reconfiguration problems that we study in this paper turn out to be computationally hard, we also look at them through the lense of parameterized complexity. As in~\cite{MouNRSS2017}, we can consider the size~$k$ of the solutions that we study (or an upper bound on them) and an upper bound~$\ell$ on the length of the reconfiguration sequence as natural parameter choices. Furthermore, we also consider neighborhood diversity as a structural parameter of the underlying graph, as started out with~\cite{GimIKO2022} in the context of reconfiguration.

In the present paper, we are going to apply the concept of reconfiguration to different notions of \emph{alliances} that have been defined in the literature, starting with \cite{FriLHHH2003,Kimetal2005,KriHedHed2004,Sha2004,SzaCza2001}. Several surveys have been written on alliances and related notions \cite{FerRod2014a,OuaSliTar2018,YerRod2017}, and even two chapters of the recent monograph \cite{HayHedHen2021} have been devoted to this topic.
Possible applications are nicely described in \cite{OuaSliTar2018}, among them also community-detection problems~\cite{SebLagKhe2012}. For instance, given a set of vertices~$A$ that should model an alliance, one could think of some $v\in A$ to be a weak spot in the alliance if it has more vertices outside of~$A$ (in a sense, enemies) in its neighborhood than allies (situated in~$A$). This idea leads to the notion of a \emph{defensive alliance}, where such weak spots are not permitted. Similarly, an \emph{offensive alliance} is longing for weak spots in the complement of~$A$ as possible points of attack. These notions will be defined more formally in the next section. However, the intuition laid so far should suffice to see that reconfiguring alliances makes a lot of sense from a practical perspective. Now, the `tokens' could be viewed as `armies' that move around, and `token sliding' would take care of the geography modeled by the underlying graph.
The main results of this paper are the following ones, where we (again) refer to the precise definitions of the problems given below.
\begin{itemize}
    \item For all variants of alliance reconfiguration problems (defensive, offensive, powerful), we can prove their \pspace-completeness for all variants of token movements.
    This remains true if the alliances are \emph{global}, i.e., if they also form dominating sets. The picture changes if we require that a (global) offensive alliance is also an independent set; then, the reachability questions are solvable in $\logspace$ and therefore much easier. For details, see \autoref{tab:Alliance-Reconfig-complexity}.
    \item We also consider different parameterizations for the (hard) reconfiguration problems. In short, all\shortversion{ but few} alliance reconfiguration problem variants\longversion{\footnote{apart from reconfiguring (independent) \offalls by token jumping}} are proven to be in \FPT\ with the combined parameter $\ell+k$, where $\ell$ upper-bounds the length of the reconfiguration sequence and $k$ denotes the number of tokens. For the powerful or global problem variations, even the parameter~$k$ alone suffices to prove membership in \FPT. Also neighborhood diversity is a nice starting point for parameterized tractability results, as we show.
    \item We introduce and discuss the novel notion of \emph{reconfiguration monotonicity} that turns out to be quite helpful in linking token addition and removal together with token jumping. These results could be interesting beyond the reconfiguration of alliances. 
\end{itemize}

\section{Definitions and Notations}
Let $\mathbb{N}$ denote the set of all nonnegative integers (including 0). For $n\in \mathbb{N}$, we will use the notation $[n]\coloneqq\{1,\ldots,n\}$. Let $G=(V,E)$ be a graph, i.e., $E\subseteq {\binom{V}{2}}$. If $X\subseteq V$, then $G[X]$ denotes the subgraph induced by~$X$, i.e., $G[X] \coloneqq (X,\{e\in E\mid e\subseteq X\})$.
$N_G(v)$ describes the \emph{open neighborhood} of $v\in V$ with respect to~$G$. The \emph{closed neighborhood} of $v\in V$  with respect to~$G$ is defined by $N[v] \coloneqq N(v)\cup \{ v \}$. For a set $A\subseteq V$, its open neighborhood is defined as $N_G(A) \coloneqq \bigcup_{v\in A} N_G(v)$. The closed neighborhood of~$A$ is given by $N_G[A]\coloneqq N_G(A) \cup A$. The \emph{degree} of a vertex $v\in V$ with respect to~$G$ is denoted $d_G(v) \coloneqq \vert N_G(v)\vert$. The \emph{boundary} of $A\subseteq V$ is defined by $\partial A\coloneqq N_G(A) \setminus A$. With $N_A(v)$ and $ d_A(v)$, we describe the open neighborhood and the degree of $v$ with respect to $G[A \cup \{v\}]$. We suppress the index~$G$ if clear from context. A vertex of degree one is called a \emph{leaf}, a vertex of degree zero is an \emph{isolate}. Similarly, an edge connecting two leaves is called an \emph{isolated edge}. A set $C\subseteq V$ is a \emph{clique} in~$G$ if $C\subseteq N_G[v]$ for each $v\in C$. A vertex~$v$ of~$G$ is called \emph{simplicial} in~$G$ if $N_G(v)$ is a clique in~$G$. For instance, leaves are always simplicial.
The ordering $v_1,\dots, v_n$ of the vertices of $G$ is a \emph{perfect elimination order} of~$G$ if, for all $i\in [n]$, $v_i$ is simplicial in $G[\{v_1,\dots,v_i\}]$.
A graph is \emph{chordal} if it has a perfect elimination ordering. The \emph{neighborhood diversity} $\nd(G)$ of a graph~$G$ is defined as the number of equivalence class of the following equivalence relation: vertices $v,u \in V$ are equivalent \iffl $N(u)\setminus \{v\} = N(v)\setminus \{u\}$. We also say $u$ and $v$ \emph{have the same type}.
 
Let $A,B \subseteq V$. Then, $A$ can be transformed to $B$ by a \emph{token removal} step if $A\subseteq B$ and $\vert B\setminus  A\vert = 1$. In this case, $B$ can be transformed by a \emph{token addition} step to~$A$. We say that $A$ can be transformed to $B$ by a \emph{token jumping} step if $\vert A\vert = \vert B \vert$, $\vert A \setminus B \vert = 1$. For $v\in A \setminus B$ and $u \in B \setminus A$, we say the token jumps from~$v$ to~$u$. A token jumping step is called a \emph{token sliding} if the vertices in $v\in A\setminus B$ and $u \in B\setminus A$ are neighbors. In this case, we say the token slides from~$v$ to~$u$. A sequence $A_1,\ldots, A_{\ell}$ is a  \emph{token addition removal sequence} (or \emph{token jumping sequence} or \emph{token sliding sequence}, respectively) if for each $i\in [\ell-1]$, $A_i$ can transformed to $A_{i+1}$ by a token addition or removal (or token jumping or token sliding, respectively) step. We also employ the abbreviations TAR
(or TJ or TS, respectively) sequence. For $Y\in \{\text{TAR},\text{TJ},\text{TS}\}$, such a sequence is named an $X$-$Y$ (reconfiguration) sequence 
if all sets~$A_i$ in this $Y$ sequence
satisfy the property $X$.

Let $G=(V,E)$ be a graph. A set $I\subseteq V$ is \emph{independent} if $G[I]$ contains only isolates. Graph~$G$ is \emph{bipartite} if $V$ can be partitioned into two independent sets. 
A set $D\subseteq V$ is called \emph{dominating} if $N[D]=V$. 
A set $A\subseteq V$ is called a \emph{defensive alliance} \shortversion{(or DA for short)} if $d_A(v) + 1 \geq d_{V \setminus A}(v)$ for each $v\in A$. 
$A \subseteq V$ with $d_A(v) \geq d_{V \setminus A}(v) + 1$ for each $v\in \partial A$ is called an \emph{offensive alliance}  \shortversion{(or OA for short)}. If a vertex set is a defensive and an offensive alliance, it is called a \emph{powerful alliance} \shortversion{(or PA for short)}. For $X\in\{$\,defensive, offensive, powerful\,$\}$, a \emph{global} $X$ alliance is an $X$ alliance that is also a dominating set. Similarly, an \offall which is also an independent set is called an \emph{independent \offall}; see~\cite{RodSig2006}.
We will now define the decision problems studied in this paper. We differentiate between two versions of reconfiguration problems. We will use $X$ as any alliance version, viewed as a property of vertex sets and abbreviated as \longversion{$\textsc{Def}$, \textsc{Off}, \textsc{Pow}}\shortversion{\textsc{D, O, P}} and sometimes prefixed with $\textsc{G}$ (global) or $\textsc{Idp}$, while $Y\in \{\text{TAR},\text{TJ},\text{TS}\}$.

\noindent
\centerline{\fbox{\begin{minipage}{.96\textwidth}
\textbf{Problem name: }\textsc{$X$-Alliance Reconfiguration}-$Y$, or  \textsc{$X$\longversion{-All}\shortversion{A}-Reconf}-$Y$ for short.\\
\textbf{Given: } A graph $G=(V,E)$ and $X$ alliances $A_s,A_t \subseteq V$ (and $k \in \mathbb{N}$ if $Y=\text{TAR}$).\\
\textbf{Question: } Is there an $X$-$Y$ reconfiguration sequence $(A_s=A_1,\ldots, A_{\ell}=A_t)$  (with $\vert A_i\vert \leq k$ for $i \in [\ell]$ if $Y=\text{TAR}$)?
\end{minipage}
}}

\noindent
\centerline{\fbox{\begin{minipage}{.96\textwidth}
\textbf{Problem name: }\textsc{Timed $X$-Alliance Reconfiguration}-$Y$, or  \textsc{T-$X$\longversion{-All}\shortversion{A}-Reconf}-$Y$ for short.\\
\textbf{Given: } A graph $G=(V,{E})$,  $X$ alliances $A_s,A_t \subseteq V$ and $T\in \mathbb{N}$ ($k \in \mathbb{N}$ if $Y=\text{TAR}$).\\
\textbf{Question: } Is there an $\ell \in \mathbb{N}$ with $\ell < T$ and an $X$-$Y$ reconfiguration sequence $(A_s=A_1,\ldots, A_{\ell}=A_t)$ (with $\vert A_i\vert \leq k$ if $Y=\text{TAR}$)?
\end{minipage}
}}
In these problems, we call $A_s$ the start configuration and $A_t$ the target configuration.
The first version asks if there is a reconfiguration sequence between $A_s$ and $A_t$, while the timed version also gives an upper bound on the number of reconfiguration steps. Sometimes, we also speak of the \emph{underlying combinatorial problem}, referring to:  given a graph $G$ and $k\in \mathbb{N}$; is there a set $D$, $|D|\leq k$, with property $X$? 
\shortversion{As usual, we will use the star symbol $(*)$ to label theorems etc. whose proof is moved to the appendix due to space constraints.}

\longversion{\paragraph*{Organization of the Paper.} In \autoref{sec:classical-complexity}, we look into classical complexity results for our problems; we find two (separating) classes: \pspace-completeness and \logspace. The hardness results motivate us to look further into aspects of parameterized complexity, focussing on the parameters `solution size' $k$ and reconfiguration length~$\ell$ in \autoref{sec:alliance-fpt} and on the parameter `neighborhood diversity' (combined with others) in \autoref{sec:alliance-nb-diversity}. We revisit our results in a concluding section, also pointing to some open problems.}

\section{\pspace-completeness or Membership in \logspace}
\label{sec:classical-complexity}

To motivate our \longversion{later} parameterized studies, we will prove \pspace-completeness for (most of) the alliance reconfiguration problems. These results are not that surprising as there are other \pspace-complete reconfiguration problems for which the underlying combinatorial problem is \NP-complete. \longversion{The problem called }\mbox{(\textsc{Timed}-)}\DSReconf-$\text{TJ}$ is such an example that will be important for us and is hence presented next.

\noindent
\centerline{\fbox{\begin{minipage}{.96\textwidth}
\textbf{Problem name: }\textsc{(Timed-)Dominating Set Reconfiguration}-token jumping, or  \textsc{(T-)DS-Reconf}-$\text{TJ}$ for short.\\
\textbf{Given: } A graph $G=(V,{E})$ and dominating sets $D_s,D_t \subseteq V$ (and $T\in \mathbb{N}$).\\
\textbf{Question: } Is there an $\ell \in \mathbb{N}$ (with $\ell < T$ and a) dominating set token jumping reconfiguration sequence $(D_s=D_1,\ldots, D_{\ell}=D_t)$?
\end{minipage}}}

We will use this problem to prove the claimed \pspace-completeness of our problems (see~\cite{BonDorOuv2021}). All the hardness proofs have the same idea: we take a \DSReconf-TJ instance $(G=(V,E), D_s, D_t)$; we construct a new graph $\widetilde{G} = (\widetilde{V}, \widetilde{E})$ with some copies $V_1,\ldots, V_p$ ($p$ depends on the alliance version that we consider) of the vertex set $V$, i.e., $V_i=\{v_i\mid v\in V\}$, and some additional vertices. If we consider the timed variant, then the very same time bound~$T$ can be also taken for the alliance reconfiguration problem.
In every case, the tokens in $V_1$ represent the tokens in the \DSReconf-TJ instance. To achieve this, we define, for each $D \subseteq V$, a set $A_D\subseteq \widetilde{V}$, such that for two sets $D,D' \subseteq V$, $A_D\setminus A_{D'}= \{ v_1\mid v\in D\setminus D'\}\subseteq V_1$. Furthermore, $D$ is a dominating set of~$G$ \iffl $A_D$ is an alliance with the right properties of~$\widetilde{G}$. This \longversion{already }implies one direction of the equivalence. For the other direction, we show that in an alliance reconfiguration sequence $A_{D_s}=A_1,\ldots, A_{\ell}=A_{D_t}$, there \longversion{exists}\shortversion{is} a $D_i$ with $A_i=A_{D_i}$ for each $i\in [\ell]$. As the reconfiguration sequences on $G$ and $\widetilde{G}$ will have the same length, we will show directly the \pspace-completeness for the timed versions that are hence not explicitly stated. 
\shortversion{We will show a detailed proof only in one of the many cases that implement this idea more concretely.}
\shortversion{Membership in \pspace immediately follows by guessing the movements in the implicit reconfiguration graph.}
\longversion{For the \pspace-membership of each of the \textsc{Alliance Reconfiguration} versions, we will describe a non-deterministic Turing machine that runs in polynomial space. At the beginning, we write both alliances on the tape. In each further step, we guess which token will be moved to which node (or in TAR: which token will be removed or where we place a token) and check if the obtained vertex set satisfies the corresponding alliance condition or if it is the target configuration. In the case that we reached the target configuration, we can return true. After $2^n$ steps, we will stop, as there are at most $2^n$ many  vertex sets and we would otherwise visit sets which we already reached before.}  

\subsection{\pspace-completeness of Token Sliding \shortversion{and Jumping}}

\begin{figure}[bt]
    \centering
    	
\begin{subfigure}[t]{.51\textwidth}
    \centering
    	
	\begin{tikzpicture}[transform shape]
		      \tikzset{every node/.style={ fill = white,circle,minimum size=0.3cm}}
            \node [rectangle,label={center:$m_{v,j,1}$}] at (0.5,-0.4){};
            \node [rectangle,label={center:$m_{v,j,2}$}] at (-1,-0.4){};
            \node [rectangle,label={center:$m_{v,j,3}$}] at (-2.7,0){};
            \node [rectangle,label={center:$m_{v,j,4}$}] at (-0,-1.4){};
            \node [rectangle,label={center:$m_{v,j,5}$}] at (0.7,1.4){};
            \node [rectangle,label={center:$m_{v,j,6}$}] at (-0.7,1.4){};
            \node [rectangle,label={center:$v_t$}] at (1.4,0){};

			\node[draw] (m1) at (0,0) {};
			\node[draw] (m2) at (-1,0) {};
			\node[draw] (m3) at (-2,0) {};
			\node[draw] (m4) at (0,-1) {};
			\node[draw] (m5) at (0.5,1) {};
			\node[draw] (m6) at (-0.5,1) {};
			\node[draw] (vt) at (1,0) {};			
            \path (m1) edge[-] (vt);		
            \path (m1) edge[-] (m2);	
            \path (m1) edge[-] (m4);	
            \path (m1) edge[-] (m5);	
            \path (m1) edge[-] (m6);	
            \path (m2) edge[-] (m3);
        \end{tikzpicture}

    \subcaption{Construction of $M_v$ ($v\in V$, $j\in [d_G(v)]$). Note that $t=3$ if $j=d_G(v)$ and $t=2$, else.}
    \label{fig:defall_TS_Space_gadget}
\end{subfigure}\qquad
\begin{subfigure}[t]{.43\textwidth}
    \centering
    	
	\begin{tikzpicture}[transform shape]
			\tikzset{every node/.style={ fill = white,circle,minimum size=0.3cm}}
            \node  at (-0.5*1.5,2.25)[rectangle]{$V_1\cup V_3\setminus\{v_1,v_3\}$};
            \node [rectangle] at (-0.5*1.5,1.75){$=$};
            \node  at (2.25*1.3,0)[rectangle,label={[rotate=-90]center:$\{u_2\mid u\in N[v]\}$}]{};
            \node [] at (1.75*1.3
            ,0){$=$};
            \node at (-0.5*1.5,1) {\dots};	
            \node at (1*1.5,0.15) {\vdots};		
            \draw (1*1.5,0) ellipse (8*1.5pt and 24pt);	
            \draw (-0.5*1.5,1) ellipse (24*1.3pt and 10pt);

			\node[draw,label={below:$v_1$}] (v1) at (0*1.5,0) {};
			\node[draw,label={below:$v_3$}] (v3) at (-1*1.5,0) {};
			\node[draw] (v11) at (0*1.5,1) {};
			\node[draw] (v12) at (-1*1.5,1) {};
			\node[draw] (v21) at (1*1.5,-0.5) {};
			\node[draw] (v22) at (1*1.5,0.5) {};
            \path (v1) edge[-] (v3);			
            \path (v1) edge[-] (v21);			
            \path (v1) edge[-] (v22);			
            \path (v1) edge[-] (v11);			
            \path (v1) edge[-] (v12);		
            \path (v3) edge[-] (v11);			
            \path (v3) edge[-] (v12);		
            
        \end{tikzpicture}

    \subcaption{Construction for $\widetilde{G}[V_1\cup V_2\cup V_3]$ ($v\in V$).}
    \label{fig:defall_TS_PSpace_consruction}
    \end{subfigure}

    \caption{Construction for \autoref{thm:defall_TS_PSpace}}
    
\end{figure}
\begin{theorem}\label{thm:defall_TS_PSpace}
\longversion{The problem }\defallianceReconf-TS is \pspace-complete, even on chordal graphs.
\end{theorem}

\begin{pf}
For \pspace-hardness, we show a reduction from \DSReconf-TJ. Hence, let $G=(V,E)$ be a graph and $D_s,D_t \subseteq V$ be dominating sets of~$G$ with $k\coloneqq \vert D_s\vert = \vert D_t\vert$. Define $\widetilde{G}=(\widetilde{V},\widetilde{E})$ with $V_q \coloneqq \{v_q \mid v \in V\}$ for $q\in [3]$ and $M_{v,p} \coloneqq \{m_{v,j,p}\mid j\in [d_G(v)]\}$ for $p \in [6]$ and $v\in V$. To simplify the notation, denote $M_{v,-} \coloneqq \bigcup_{p=1}^{6} M_{v,p}$ for $v\in V$ as well as $M_{-,p} \coloneqq \bigcup_{v\in V} M_{v,p}$ for $p \in [6]$. \longversion{Let}
    \begin{equation*}
        \begin{split}
            E_M\coloneqq{}& \{\{m_{v,j,1},m_{v,j,p}\}, \{m_{v,j,2},m_{v,j,3}\} \mid v\in V, \, j\in [d_G(v)], p\in\{2,4,5,6\} \}\,,\\  
            \widetilde{V}\coloneqq{}& \left(\bigcup_{q=1}^3 V_q\right) \cup \left(\bigcup_{v\in V} M_{v,-}\right)\,,\text{ and}\\
            \widetilde{E}\coloneqq{}&   \binom{V_1 \cup V_3}{2} \cup E_M \cup \{ \{v_1,u_2\} \mid v,u\in V,\, u\in N_G[v]\}  \cup{} \\
            & \{\{v_2 ,m_{v,j,1}\} \mid v\in V, j\in [d_G(v)-1]\} \cup \{\{v_3,m_{v,d_G(v),1}\}\mid v\in V\}\,.
        \end{split}
    \end{equation*}

     $\tilde{G}$ is chordal as there is a perfect elimination ordering. (1)  The vertices in $M_{-,3}\cup M_{-,4} \cup M_{-,5} \cup M_{-,6}$ are leaves. (2) After deleting these vertices, $M_{-,2}$ are leaves. (3) On $\widetilde{G}[V_1\cup V_2 \cup V_3 \cup M_{-,1}]$, the vertices in $M_{-,1}$ are leaves. (4) $V_2$ are  simplicial vertices on $\widetilde{G}[V_1\cup V_2 \cup V_3]$ and (5) $V_1\cup V_3$ is a clique. 

    \noindent
    For $D \subseteq V$, define $D' \coloneqq \{v_1 \mid v\in D\}$ and $A_D\coloneqq D' \cup  V_2 \cup V_3  \cup M_{-,1} \cup M_{-,2} \subseteq \widetilde{V}$. 
    \begin{claim}\label{claim:DS_to_DefAlliance}\shortversion{$(\star)$}
        Let $D\subseteq V$. $D$ is a dominating set of~$G$ \iffl $A_D$ is a \defall of~$\widetilde{G}$.
    \end{claim}
    \begin{toappendix}
    \begin{pfclaim}\shortversion{[of \autoref{claim:DS_to_DefAlliance}]} 
        If $D$ is empty, then $D$ is not a dominating set. Furthermore, $A_D$ is not a \defall, as for each $v_2\in V_2$, $\vert N_{A_D}(v_2)\vert + 1 = d_G(v) < d_G(v) + 1 = \vert N_{\widetilde{V} \setminus A_D}(v_2)\vert $. Therefore, we can assume that $D\neq \emptyset$. 
        Hence, for each $v\in V$, $N_{\widetilde{V}\setminus A_D}(v_3) \subsetneq V_1 \cup \{m_{v,d_G(v),1}\}$. Thus, $d_{\widetilde{V}\setminus A_D }(v_3) \leq \vert V\vert < \vert V\vert + 1 \leq \vert (V_3\setminus\{v_3\})\cup D' \cup \{m_{v,d_G(v),1}\}\vert + 1 = d_{A_D}(v_3) + 1$.
        For $v_1 \in D'$,  $N_{\widetilde{V} \setminus A_D}(v_1)\subsetneq V_1$ and $V_3\subsetneq N_{A}(v_1)$ imply $d_{\widetilde{V} \setminus A_D}(v_1)\leq d_{A_D}(v_1) + 1$.
        Let $v\in V$, $j\in [d_G(v)]$. $m_{v,j,1}$ has, besides $m_{v,j,2}$, one more neighbor in $A_D$ (namely, $v_3$ if $j=d_G(v)$  and $v_2$, otherwise). Thus, $d_{A_D}(m_{v,j,1}) + 1 = 3 = d_{\widetilde{V}\setminus A_D}(m_{v,j,1})$. Furthermore, $d_{A_D}(m_{v,j,2}) + 1 = 2 > 1 = d_{\widetilde{V}\setminus A_D}(m_{v,j,2})$.
        This leaves us to show that, for each $v\in V$, $d_{\widetilde{V} \setminus A_D}(v_2) \leq d_{A_D}(v_2) + 1$ \iffl $D$ is a dominating set.  
        \\
        ``$\Leftarrow$'': Let $D$ be a dominating set of~$G$. Then for each $v_2\in  V_2$, there exists a $u_1\in D' \cap N(v_2)$. This implies $\{m_{v,j,1} \mid j \in [d_{G}(v)-1]\}\cup \{u_1\} \subseteq N_{A_D}(v_2) $ and $ N_{\widetilde{V} \setminus A_D}(v_2)\subsetneq \{w_1\in V_1 \mid w\in N_G[v]\}$. Therefore, $d_{\widetilde{V}\setminus A_D}(v_2) \leq d_{G}(v) < d_G(v) + 1 \leq d_{A_D}(v_2) + 1.$
\longversion{\\} ``$\Rightarrow$'':
        If $D$ is not a dominating set, then there is a  $v_2\in V_2$ such that $N_G[v]\cap D=\emptyset$. Thus, $N_{D'}(v_2)=\emptyset$ and $d_{A_D}(v_2) + 1  = d_G(v) < d_G(v)+1  =d_{\widetilde{V}\setminus A_D}(v_2)$. Hence, $A_D$ is not a \defall.  
    \end{pfclaim}
    \end{toappendix}
     This\longversion{ claim directly} shows that $(G,D_s,D_t)$ is a \yes-instance of \DSReconf-TJ only if $(\tilde{G},A_{D_s},A_{D_t})$ is a \yes-instance of \defallianceReconf-TS. To see this, we transform each dominating set $D_i$ in our sequence into the \defall~$A_{D_i}$. 

     Now assume there exists a \defalltslideseq $A_{D_s}=A_1,\ldots,A_{\ell}=A_{D_t}$.  If we can show that for each $i\in [\ell]$, $V_2 \cup V_3 \cup  M_{-,1} \cup M_{-,2} = A_i \setminus V_1$, then the claim implies that there exist dominating sets $D_s=D_1,\ldots, D_{\ell}=D_t$ with $A_{D_i}=A_i$ for $i\in [\ell]$, so that  $D_1,\ldots, D_{\ell}$ is a \DSReconf-TJ sequence.

      We will show this by contradiction. To this end, let $i\in [\ell]$ ($i\neq 1$) be the first index such that $V_2 \cup V_3 \cup  M_{-,1} \cup M_{-,2} \neq A_i \setminus V_1$.
      Let $v\in V$ and $ j\in [d_G(v)]$. Assume $m_{v,j,1} \notin A_i$. As we consider token sliding, there exists a $p \in \{4,5,6\}$ with $m_{v,j,p}\in A_i$. Then $d_{A_i}(m_{v,j,2}) + 1 = 1 < 2 = d_{\widetilde{V} \setminus A_i}(m_{v,j,2})$. Therefore, $M_{-,1} \subseteq A_i$ and as $m_{v,j,1}$ is the only neighbor of $m_{v,j,4},m_{v,j,5},m_{v,j,6}$, $(M_{-,4} \cup M_{-,5} \cup M_{-,6}) \cap A_i=\emptyset$. Since $m_{v,j,3}$ is the only neighbor of $m_{v,j,2}$ in $\widetilde{V} \setminus A_ {i-1}$ and $m_{v,j,2}$ is the only one of $m_{v,j,3}$ in $A_{i-1}$, $m_{v,j,3}\in A_i$ \iffl $m_{v,j,2}\notin A_i$. In this case, $d_{A_i}(m_{v,j,1}) +1 \leq2 <4= d_{\widetilde{V} \setminus A_i}(m_{v,j,1})$. Thus, $\left(\bigcup_{p=1}^{6} M_{-,p}\right)\cap A_i = M_{-,1} \cup M_{-,2}$.  
      If $v_3\notin A_i$ for some $v\in V$, $d_{A_i}(m_{v,d_G(v),1}) +  1 = 2 < 4 = d_{\widetilde{V} \setminus A_i}(m_{v,d_G(v),1})$, since $m_{v,d_{G}(v),4}, m_{v,d_{G}(v),5}, m_{v,d_{G}(v),6} \notin N_{\widetilde{G}}(v_3)$. Thus, $V_3\subseteq A_i$. Analogously, $V_2\subseteq A_i$.  Hence,  $V_2 \cup V_3 \cup M_{-,1} \cup M_{-,2}= A_i \setminus V_1$.
\end{pf}

Observe that each $A_{D_i}$ is also a dominating set\longversion{ of~$\widetilde{G}$}: For $i\in [\ell]$, \longversion{the only vertices in $\widetilde{V} \setminus A_{D_i}$ are in $V_1\cup M_{-,3} \cup M_{-4}\cup M_{-,5}\cup M_{-,6}\subseteq N(V_3 \cup M_{-,1} \cup M_{-,2})\subseteq N(A_{D_i})$}\shortversion{ $\widetilde{V} \setminus A_{D_i}\subseteq V_1\cup M_{-,3} \cup M_{-4}\cup M_{-,5}\cup M_{-,6}\subseteq N(V_3 \cup M_{-,1} \cup M_{-,2})\subseteq N(A_{D_i})$}. 
For \longversion{the vertices }$v_1\in V_1$, we also know $d_{\widetilde{V} \setminus A_{D_i}}(v_1) + 1 \leq \vert V_1\vert = \vert V_3 \vert \leq d_{A_{D_i}}(v_1)$, as $A_{D_i}\neq \emptyset$. Furthermore, $d_{\widetilde{V} \setminus A_{D_i}}(m_{v,j,p}) + 1 = 1 = d_{A_{D_i}}(m_{v,j,p})$. 
Hence, the\longversion{ sets} $A_{D_i}$ are even global \powalls. 

\begin{corollary}\label{cor:global-TS}\longversion{The problems}
\textsc{G-}\defallianceReconf-TS, \longversion{\textsc{Pow}-\AllReconf}\shortversion{\textsc{PA-Reconf}}-TS, \longversion{as well as \textsc{G-Pow}-\AllReconf}\shortversion{\textsc{G-PA-Reconf}}-TS are \pspace-complete, even on chordal graphs.
\end{corollary}

Even if $A_{D_i}$ is an \offall, this construction does not provide a proof for the \pspace-hardness of \offallianceReconf-TS, as the \defall property is necessary\longversion{ for this construction to work}. Namely, we could move the tokens in $V_1$ as we want\longversion{ (if the remaining tokens stay at their vertices)}. Thus,  $(\widetilde{G},A_{D_s},A_{D_t})$  is always a \yes-instance as an \offallianceReconf-TS instance. Even if we bound the number of steps, this is a \yes-instance \iffl $\vert A_{D_s}\setminus A_{D_t} \vert\leq \ell$. \longversion{Hence, we need a new yet similar construction \longversion{for  the \pspace-completeness of \offallianceReconf-TS}\shortversion{for \textsc{Off}}.}


\shortversion{
With quite similar reductions, yet displaying non-trivial adaptations to the specific cases, we can show the following list of complexity results for \longversion{token jumping and sliding}\shortversion{TJ and~TS}.
\begin{theorem}\label{thm:all_All_TJ}$(*)$ For any \longversion{$X\in\{$\,\textsc{\,Def}, \textsc{ G-def}, \textsc{ Off}, \textsc{ G-Off}, \textsc{ Idp-Off}, \textsc{ Pow}, \textsc{ G-Pow}\,$\}$, $X$-\AllReconf-TJ}\shortversion{$X\in\{\textsc{DA,G-DA,OA,G-DOA,Idp-OA,PA,G-PA}\}$, $X$-\textsc{Reconf}-TJ} is 
\pspace-complete, even on bipartite graphs. Moreover,
      \offallianceReconf-TS, \textsc{G-}\offallianceReconf-TS  and \textsc{G-}\offallianceReconf-TJ are \pspace-complete, even on chordal graphs.  
\end{theorem}
}     
\begin{toappendix}

\shortversion{We will proof \autoref{thm:all_All_TJ} by showing the \pspace-completeness for each problem itself.}\begin{figure}[bt]
\centering
    	
\begin{subfigure}[t]{.51\textwidth}
    \centering
    	
	\begin{tikzpicture}[transform shape]
		      \tikzset{every node/.style={ fill = white,circle,minimum size=0.3cm}}
            \node  at (-2,0)[label={[rectangle,rotate=90]above:$\{u_1\mid u\in N[v]\}$}]{};
            \node at (0.7,1) {\ldots};\node () at (1.5,1.7) {$m_{v,d(v)+1}$};
            \node [label={above:$=$}] at (-1.8,-0.5){};

			\node[draw,label={below:$v_{12}$}] (v12) at (0.5,0) {};
			\node[draw,label={above:$m_{v,1}$}] (m1) at (0,1) {};
			\node[draw,] (m2) at (1.4,1) {};
			\node[draw,label={below:$v_{13}$}] (v13) at (1.5,0) {};
			\node[draw,label={below:$v_{14}$}] (v14) at (2.5,0) {};
			\node[draw] (v1) at (-1,0.75) {};
			\node[draw] (v'1) at (-1,-0.75) {};
            \node at (-1,0.15) {\vdots};
            \draw (-1,0) ellipse (15pt and 35pt);			
            \path (m1) edge[-] (v12);
			\path (m2) edge[-] (v12);			
            \path (v1) edge[-] (v12);
			\path (v'1) edge[-] (v12);
			\path (v13) edge[-] (v12);
			\path (v14) edge[-] (v13);
        \end{tikzpicture}

    \subcaption{Gadget that verifies that each vertex in the original graph is dominated ($v\in V$).}
    \label{fig:offall_TS_Space_gadget}
\end{subfigure}\qquad
\begin{subfigure}[t]{.43\textwidth}
    \centering
    	
	\begin{tikzpicture}[transform shape]
			\tikzset{every node/.style={ fill = white,circle,minimum size=0.3cm}}

			\node[draw,label={above:$v_{j}$}] (vj) at (0,0) {};
			\node[draw,label={above:$v_{j+1}$}] (vj1) at (1,0) {};
			\node[draw,label={below:$v_{j+2}$}] (vj2) at (1,-1) {};
			\node[draw,label={above:$v_{j+3}$}] (vj3) at (2,0) {};
			\node[draw,label={above:$v_{j+4}$}] (vj4) at (3,0) {};		
            \path (vj) edge[-] (vj1);
			\path (vj1) edge[-] (vj2);			
            \path (vj1) edge[-] (vj3);
			\path (vj3) edge[-] (vj4);
        \end{tikzpicture}

    \subcaption{Construction to ensure that $v_2$ stays in the alliance ($j\in \{2,7\}$, $v\in V$).}
    \label{fig:ofall_TS_PSpace_consruction}
    \end{subfigure}

    \caption{Construction for \autoref{thm:offall_TS_PSpace}}
\end{figure}   

\begin{theorem}\shortversion{$(*)$}\label{thm:offall_TS_PSpace}
 \longversion{The problem}   \offallianceReconf-TS is \pspace-complete, even on chordal graphs.
\end{theorem}

\begin{pf}\shortversion{[of \autoref{thm:offall_TS_PSpace}]}
    For the hardness part, we will use again a reduction from \DSReconf-TJ. Therefore, let $G=(V, E)$ be a graph and $D_s,D_t \subseteq V$ be dominating sets with $k\coloneqq\vert D_s\vert =\vert D_t\vert$. Define $\widetilde{G} \coloneqq (\widetilde{V},\widetilde{E})$ with $V_q\coloneqq  \{v_q\mid v\in V \}$ for $q\in  [14]$, $  M_v\coloneqq  \{ m_{v,j} \mid j\in [d_G(v)+1]\}$ for  $v\in  V$,
    \begin{equation*}
        \begin{split}
            \widetilde{V} \coloneqq & \left(\bigcup^{14}_{q=1} V_q \right) \cup \left(\bigcup_{v\in V} M_v \right)\,, \\
            \widetilde{E} \coloneqq & \binom{V_1 \cup V_2\cup V_7}{2}\cup \{\{v_1, u_{12}\}\mid v,u\in V,\, u\in N[v] \} \\
            & \cup\{\{v_q,v_{q+1}\}, \{v_{q+1},v_{q+2}\}, \{v_{q+1},v_{q+3}\}, \{v_{q+3},v_{q+4}\}\mid v\in V, q\in \{2,7\}\}\\ 
            & \cup \{\{v_{12},v_{13}\}, \{v_{13},v_{14}\},\{v_{12}, m_{v, j}\}\mid v\in V, j\in [d_G(v)+1]\} .
        \end{split}
    \end{equation*}
    $\widetilde{G}$ is a chordal graph: The vertices in $B_1 \coloneqq V_4\cup V_6 \cup V_9 \cup V_{11} \cup V_{14} \cup \left(\bigcup_{v\in V} M_v \right) $ are leaves. After removing $B_1$ from~$\widetilde{G}$, $B_2 \coloneqq V_{5} \cup V_{10}\cup V_{13}$ includes only leaves. Now, $ V_3 \cup V_8 \cup V_{12}$ is the set of simplical vertices in $\widetilde{G}[\widetilde{V}\setminus (B_1 \cup B_2)]$ and $V_1 \cup V_2 \cup V_7$ is a clique. Thus, there is a perfect elimination order on $\widetilde{V}$.
    Furthermore, we define for a vertex set $D\subseteq V$ a set 
    $$A_D \coloneqq \{v_1 \mid v\in D\} \cup V_2\cup V_4 \cup V_7 \cup V_9 \cup \left(\bigcup_{v\in V} M_v \right).$$ 
    \begin{claim}\label{claim:off_all_ts_pspace}
        Let $D \subseteq V$. $D$ is a dominating set of~$G$ \iffl $A_D$ is a \offall of~$\widetilde{G}$. 
    \end{claim}
    
\begin{pfclaim}
For the proof, we define $A\coloneqq A_D$ and $A' \coloneqq V_2\cup V_4 \cup V_7 \cup V_9 \cup \left(\bigcup_{v\in V} M_v \right)$, i.e., $A'=A\setminus V_1$. Clearly, $\partial A \subseteq V_1\cup V_3 \cup V_8 \cup V_{12} $. For all $v\in V$ and $q\in\{3,8\}$, we know  $d_A(v_q) = \vert \{ v_{q-1},v_{q+1}\} \vert = 2 > 1 = \vert \{v_{q+2}\}\vert= d_{\widetilde{V} \setminus A}(v_q).$ Let $v\in V$. Then $N_A(v_1) = V_2 \cup V_7$ and  $N_{\widetilde{V}\setminus A}(v_1) \subseteq ( V_1\setminus\{v_1\}) \cup V_{12}$. Therefore, $d_{\widetilde{V}\setminus A}(v_1) \leq 2\,\vert V\vert - 1 <  2\,\vert V\vert = d_A(v_1)$. 
        
This leaves us to check $V_{12}$.
Let $D$ be a dominating set of~$G$.
Hence, for each $v\in V$, there is some $u\in D\cap N[v]$. Therefore, for each $v_{12}\in V_{12}$, there is a $u_{1}\in N(v_{12})\cap V_1\cap A$. Hence, $N_{\widetilde{V}\setminus A} (v)\subsetneq N_{V_1}(v) \cup \{v_{13}\}$. Therefore, we have $d_{\widetilde{V}\setminus A}(v_{12}) \leq d_G(v) + 1 < d_G(v) + 2= \vert M_v\cup \{ u_1 \} \vert \leq d_{A}(v_{12})$. So, $A$ is an \offall.

Let D be no dominating set. Thus, there exists a $v\in V$ with $N[v]\cap D=\emptyset$. So $d_{\widetilde{V} \setminus A}(v_{12})= \vert \{ u_1 \mid u\in N_G[v]\}\cup \{ u_{13} \}\vert =d_G(v)+2 > d_G(v) + 1= \vert M_v\vert = d_{A}(v_{12})$ and $A$ is no \offall. 
\end{pfclaim}
The claim 
implies that $(G,D_s,D_t)$ is a \yes-instance of \DSReconf-TJ only if $(\widetilde{G},A_{D_s},A_{D_t})$ is a \yes-instance of \offallianceReconf-TS.  To see this, let $D_s=D_1,\ldots,D_{\ell} = D_t$ be a \dstjumpseq. By \autoref{claim:off_all_ts_pspace}, $A_{D_s}=A_{D_1},\ldots, A_{D_{\ell}}=A_{D_s}$ is a sequence of \offalls on $\widetilde{G}$. Since $V_1$ is a clique and only the vertices in $V_1$ move, this is an \offalltslideseq. 

For the if-part, assume we have an \offalltslideseq $A_{D_s}=A_1,\ldots, A_{\ell}= A_{D_t}$. We will inductively  show that, for each $i \in [\ell]$, $$B_i\coloneqq\left( \left( \bigcup_{q=2}^{14} V_q  \right) \cup \left(\bigcup_{v\in V} M_v \right)\right) \cap  A_i= V_2\cup V_4 \cup V_7 \cup V_9 \cup \left(\bigcup_{v\in V} M_v \right)=:C\,.$$
Together with \autoref{claim:off_all_ts_pspace}, this implies  that,  for each $i\in [\ell]$, $D_i = \{v\in V\mid v_1\in A_i \}$ is a dominating set with $\vert D_i\vert = \vert D_s\vert$. As $A_{D_i}=\{v_1\mid v \in D_i\}\cup C$, $D_s=D_1,\ldots, D_{\ell}=D_t$ is a \dstjumpseq.

Trivially, $B_1 =C$.  Assume $ B_{i-1}=C$ for $i\in[\ell] \setminus \{ 1 \}$. First, we show $B_i \subseteq C$. As $A_i$ can be reconfigured from $A_{i-1}$ by one token sliding step and $V_5 \cup V_6 \cup V_{10} \cup V_{11} \cup V_{13} \cup V_{14}$ has no neighbor in $A_{i-1}$, $A_i\cap (V_5 \cup V_6 \cup V_{10} \cup V_{11} \cup V_{13} \cup V_{14})= \emptyset$. Assume there exists a $v_q \in B_i$ for $q\in\{3,8,12\}$. Then $v_{q+2}\in \partial A_i$ with $d_{\widetilde{V} \setminus A_i}(v_{q+2})=1= d_{A_i}(v_{q+2})$. This would contradict the \offall property of~$A_i$. Thus, $B_i \subseteq C$. As $N \left( V_4 \cup V_9 \cup \left(\bigcup_{v\in V} M_v \right)\right) \subseteq V_3 \cup V_8 \cup V_{12}$, the tokens in $V_4 \cup V_9 \cup \left(\bigcup_{v\in V} M_v \right)$ will not move in this step. Assume there is a $v_2\in V_2$ with $v_2\notin B_i$. Recall $N[v_2]=V_1 \cup V_2 \cup V_7 \cup \{v_3\}$. As we have proved that $v_3\notin B_i$, $d_{A_i}(v_3)=\vert \{v_4\}\vert = 1 \leq 2 = \vert \{v_2,v_5\} \vert =d_{\widetilde{V}\setminus A_i}(v_3)$. This would contradict the  \offall property of $A_i$. Therefore, $V_2 \subseteq B_i$. The proof for $V_7\subseteq B_i$ works analogously. Hence $B_i=C$ for each $i\in [\ell]$. With the reasoning above, this completes the proof.
\end{pf}

\longversion{\subsection{\pspace-completeness of Token Jumping}\label{subsec:tj_Pspace}}

\begin{theorem}\label{thm:defall_TJ_PSpace}
 \longversion{The problem}       \defallianceReconf-TJ is \pspace-complete, even on bipartite graphs.
\end{theorem}
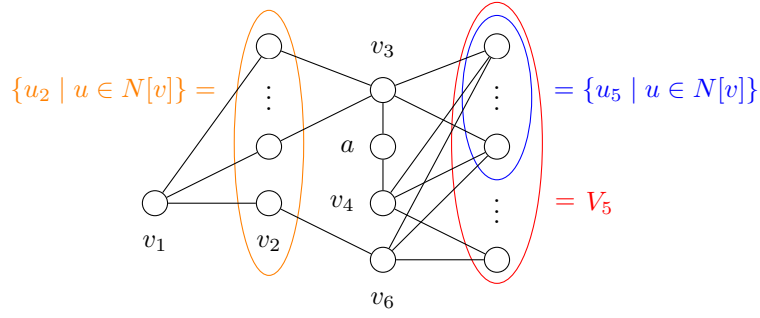
\begin{figure}[bt]
    \centering
    	
	\begin{tikzpicture}[transform shape]
		      \tikzset{every node/.style={ fill = white,circle,minimum size=0.3cm}}
            \node [rectangle]  at (3.6,0) {\color{blue}$=\{u_5\mid u\in N[v]\}$};
            \node [rectangle] at (-3.5,0)
            {\color{orange}$\{u_2\mid u\in N[v]\}={}$};

			\node[draw,label={above:$v_3$}] (v3) at (0,0) {};
			\node[draw,label={below:$v_2$}] (v2) at (-1.5,-1.5) {};
			\node[draw] (v21) at (-1.5,0.58) {};
			\node[draw] (v22) at (-1.5,-0.75) {};
			\node[draw] (v51) at (1.5,0.58) {};
			\node[draw] (v52) at (1.5,-0.75) {};
			\node[draw] (v53) at (1.5,-2.25) {};

			\node[draw,label={left:$a$}] (a) at (0,-0.75) {};
			\node[draw,label={left:$v_4$}] (v4) at (0,-1.5) {};
			\node[draw,label={below:$v_6$}] (v6) at (0,-2.25) {};
			\node[draw,label={below:$v_1$}] (v1) at (-3,-1.5) {};
   
            \node at (1.5,0.0) {\vdots};
            \node at (-1.5,0.0) {\vdots};
            \node at (1.5,-1.5) {\vdots};
            \node at (2.85,-1.5) {\color{red}$V_5$};
            \node at (2.4,-1.5){\color{red}$=$};
            \draw [blue] (1.5,-0.1) ellipse (13pt and 31pt);
            \draw [red] (1.5,-0.7) ellipse (17pt and 53pt);
            \draw [orange](-1.5,-0.7) ellipse (13pt and 50pt);			
            \path (v21) edge[-] (v3);
			\path (v22) edge[-] (v3);		
            \path (v51) edge[-] (v3);
			\path (v52) edge[-] (v3);		
            \path (v51) edge[-] (v4);
			\path (v52) edge[-] (v4);
			\path (v53) edge[-] (v4);
            \path (v51) edge[-] (v6);
			\path (v52) edge[-] (v6);
			\path (v53) edge[-] (v6);
			\path (v2) edge[-] (v1);
			\path (v2) edge[-] (v6);
			\path (v21) edge[-] (v1);
			\path (v22) edge[-] (v1);
			\path (a) edge[-] (v3);
			\path (v4) edge[-] (a);
        \end{tikzpicture}

        \label{fig:defall_TJ_PSpace}
    \caption{Construction of \autoref{thm:defall_TJ_PSpace} for $v\in V$.}
\end{figure}
\begin{pf}
Again, we will reduce from \textsc{(T-)DS-Reconf}-TJ. Let $G=(V,E)$ be a graph without isolates and let $D_s,D_t \subseteq V$ be two dominating sets with $k \coloneqq \vert D_s\vert = \vert D_t \vert$. 

Define the graph $\widetilde{G}=\left( \widetilde{V}, \widetilde{E} \right)$ with $V_q = \{v_q \mid v\in V\}$ for $q \in [6]$, $a \notin V$ and
\begin{equation*}
    \begin{split}
        \widetilde{V} \coloneqq &\, \{ a \} \cup \left( \bigcup_{q=1}^6 V_q\right),\\
        \widetilde{E} \coloneqq &\, \{\{v_1,u_2\},\{v_3,u_2\}, \{v_3,u_5\}\mid \{ v,u \}\in E\} \cup \{\{v_4,u_5\},\{v_6,u_5\}\mid v,u\in V\} \\
        &\cup \{\{v_3,a\}, \{v_4,a\}, \{v_1,v_2\}, \{v_6,v_2\} , \{v_3,v_5\} \mid v\in V\} .    
    \end{split}
\end{equation*}
This graph is bipartite with the partition $A= V_1\cup V_3\cup V_4 \cup V_6$ and $B = \{a\}\cup V_2\cup V_5$. This can be easily seen, as each edge in the definition of $\widetilde{E}$ first mentions the vertex of~$A$ and then the vertex of~$B$. For a vertex set $D\subseteq V$, we define $A_D = \{v_1\mid v\in D\} \cup V_2\cup V_3 \cup \{a\}$. 

 \begin{claim}\shortversion{$(*)$}\label{claim:def_all_tj_pspace}
     Let $D\subseteq V$. $D$ is a dominating set of~$G$ \iffl $A_D$ is a \defall of~$\widetilde{G}$.
\end{claim}
\begin{pfclaim}
    For the proof of the claim, we denote $A \coloneqq A_D$. As $d_A(a) + 1=\vert V_3 \vert + 1 \geq \vert V_4\vert =  d_{\widetilde{V} \setminus A}(a)$, we do not need to consider $a$ in the following. Furthermore, for each $v_1\in V_1$, $N_{\widetilde{G}}(v_1) \subseteq V_2\subseteq A$. For $v_3\in V_3$, $d_A(v_3)+1= d_G(v) + 2 > d_G(v) + 1 = d_{\widetilde{V} \setminus A}(v_3)$. Therefore, we only need to check $d_A(v_2) + 1 \geq d_{\widetilde{V} \setminus A}(v_2)$ for $v_2 \in V_2$.

    Assume $D$ is a dominating set. Then for each $u\in V$, $N_G[u] \cap D$ is not empty. Therefore, there exists a $v_1\in N_{\widetilde{G}}(u_2)\cap A\cap V_1$. Hence, $d_{\widetilde{V} \setminus A}(u_2) \leq \vert (N_{V_1}(u_2) \setminus \{v_1\}) \cup \{u_6\}\vert = d_G(u) + 1  < d_G(u) + 2 = \vert N_{V_3}(u_2) \cup \{v_1\} \vert + 1 \leq d_{A}(u_2) + 1$. Thus, $A$ is a \defall. 

    Assume $D$ is not a dominating set. So there exists a $u \in V$ such that $N_G[u] \cap D = \emptyset$. Thus, $N_{\widetilde{G}}(u_2)\cap A\cap V_1$ is also empty. Therefore, $d_{\widetilde{V} \setminus A}(u_2) =  d_{V_1}(u_2) + 1 = d_G(u) + 2 > d_G(u) + 1 = d_{V_3}(u_2) +1 = d_{A}(u_2) + 1$. Hence, $A$ is not a \defall. 
\end{pfclaim}  
\noindent
With the same arguments as in the proofs above, one can show: \begin{claim}\shortversion{$(*)$}\label{claim:def_all_tj_sequences} $D_s=D_1,\ldots,D_{\ell}=D_t$ is a \dstjumpseq on $G$ \iffl $A_{D_s}=A_{D_1},\ldots, A_{D_{\ell}}=A_{D_t}$ is a \defalltjumpseq of~$\widetilde{G}$. \shortversion{\qed}
\end{claim}
\begin{pfclaim}
Let $A_{D_s}= A_1, \ldots, A_{\ell} = A_{D_t}$ be a \defalltjumpseq of~$\widetilde{G}$. We will show by induction that, for each $i\in [\ell]$, there exists a dominating set $D_i \subseteq V$ such that $A_i = A_{D_i}$ and that $D_i\triangle D_{i-1} = \{v\mid v_1\in A_i\triangle A_{i-1}\}$ if $i>1$. For $A_1$, this is true. So assume that for $A_i$ there exists a dominating set $D_i$ such that $A_i=A_{D_i}$. Therefore, $V_2\cup V_3 \cup \{a\} = A_i \setminus V_1$. Let $x \in A_{i}\setminus A_{i+1}$ and $y\in A_{i+1}\setminus A_{i}$. Therefore, $x\notin V_4\cup V_5 \cup V_6$ and $y\notin V_2\cup V_3\cup  \{a\}$.  Furthermore, $y \notin V_5 \cup V_6$, as $d_{A_{i+1}}(y) + 1 < \vert V \vert \leq  d_{\widetilde{V}\setminus A_{i+1}}(y)$ would contradict the \defall property of $A_{i+1}$. Analogously, $y\notin V_4$. This leads to $y\in V_1$. 
Since $G$ includes no isolates, each vertex in $V_2$ has at least one neighbor in $V_3$. If $x=v_2\in V_2$, then for a $u_3\in N(v_2)\cap V_3$, $d_{A_{i+1}}(u_3) + 1 = d_G(u) + 1 < d_G(u) + 2 = d_{\widetilde{V} \setminus A_i+1}(u_3) $, contradicting the \defall property of $A_{i+1}$. Since $a$ is in each neighborhood of any vertex in $V_3$, the same argument implies $x\neq a$. For $x=v_3$ with $v\in V$, $d_{A_{i+1}}(a) + 1= \vert V_3\vert < \vert V_4\vert + 1 = d_{\widetilde{V} \setminus A_{i+1}} (a)$. This implies $x\in V_1$ and the existence of a set $D_i\subseteq V$ such that $A_{i} = A_{D_{i}}$ for each $i\in [l]$. By \autoref{claim:def_all_tj_pspace}, $D_i$ is a dominating set. Therefore, the existence of a \defalltjumpseq $A_{D_s}=A_1,\ldots, A_{\ell}=A_{D_t}$  implies the existence of a \dstjumpseq $D_s=D_1,\ldots,D_{\ell}=D_t$ as claimed. 
\end{pfclaim}    
\shortversion{\renewcommand{\qedsymbol}{}}
\longversion{\noindent This last claim finishes the whole proof.}
\end{pf}

\shortversion{\vspace{-5ex}\mbox{}}
As in the proof of \autoref{thm:defall_TS_PSpace}, $A_D$ is a dominating set of~$\widetilde{G}$ for any set $D \subseteq V$, so that we conclude the following result immediately.

\begin{corollary}\label{cor:globaldefall_TJ_PSpace}
\longversion{The problem}    \textsc{G-}\defallianceReconf-TJ is \pspace-complete, even on bipartite graphs.
\end{corollary}

\longversion{Again, w}\shortversion{W}e have to adapt our construction \shortversion{of \autoref{thm:defall_TJ_PSpace}} considerably to show an analogous result for \offalls.

\begin{theorem}\label{thm:offall_TJ_PSpace}
\longversion{The problem}     \offallianceReconf-TJ is \pspace-complete, even on bipartite graphs.
\end{theorem}

\begin{figure}[bt]
    \centering
    	
\begin{subfigure}[t]{.51\textwidth}
    \centering
    	
	\begin{tikzpicture}[transform shape]
		      \tikzset{every node/.style={ fill = white,circle,minimum size=0.3cm}}
            \node  at (2,0)[label={[rectangle,rotate=-90]above:$\{u_2\mid u\in N[v]\}$}]{};
            \node [label={above:$=$}] at (1.8,-0.5){};
            \node  at (-2,0)[rectangle,label={[rectangle,rotate=90]above:$\{u_1\mid u\in N[v]\}$}]{};
            \node [label={above:$=$}] at (-1.8,-0.5){};

			\node[draw,label={above:$v_3$}] (v3) at (0,0) {};
			\node[draw] (m1) at (-1,0.75) {};
			\node[draw] (m2) at (-1,-0.75) {};
			\node[draw] (v11) at (1,0.75) {};
			\node[draw] (v12) at (1,-0.75) {};
   
			\node[draw,label={below:$v_4$}] (v4) at (0,-1.5) {};
			\node[draw,label={below:$v_5$}] (v5) at (1,-1.5) {};
			\node[draw,label={below:$v_6$}] (v6) at (2,-1.5) {};
   
            \node at (1,0.15) {\vdots};
            \node at (-1,0.15) {\vdots};
            \draw (1,0) ellipse (15pt and 35pt);
            \draw (-1,0) ellipse (15pt and 35pt);			
            \path (m1) edge[-] (v3);
			\path (m2) edge[-] (v3);
			\path (v11) edge[-] (v3);
			\path (v12) edge[-] (v3);
			\path (v4) edge[-] (v3);
			\path (v4) edge[-] (v5);
			\path (v5) edge[-] (v6);
        \end{tikzpicture}

    \subcaption{Gadget that verifies that each vertex in the original graph is dominated ($v\in V$).}
    \label{fig:offall_TJ_Space_gadget}
\end{subfigure}
\begin{subfigure}[t]{.43\textwidth}
    \centering
    	
	\begin{tikzpicture}[transform shape]
			\tikzset{every node/.style={ fill = white,circle,minimum size=0.3cm}}
            
            \node[draw,label={above:$v_{2}$}] (v2) at (0,0) {};
			\node[draw,label={above:$b$}] (b) at (1,0) {};
			\node[draw,label={below:$a$}] (a) at (1,-1) {};
			\node[draw,label={above:$v_{7}$}] (v7) at (2,0) {};
			\node[draw,label={above:$v_{8}$}] (v8) at (3,0) {};
			\node[draw,label={above:$v_{9}$}] (v9) at (4,0) {};		
            \path (v2) edge[-] (b);
			\path (b) edge[-] (a);			
            \path (b) edge[-] (v7);
			\path (v7) edge[-] (v8);
			\path (v8) edge[-] (v9);
        \end{tikzpicture}

    \subcaption{Construction to ensure that $v_2$ stays in the alliance ($v\in V$).}
    \label{fig:offall_TJ_PSpace_consruction}
    \end{subfigure}

    \caption{Construction for \autoref{thm:offall_TJ_PSpace}}
\end{figure}

\begin{pf}
As before, we will use \textsc{(T-)DS-Reconf}-TJ for the \pspace-hardness proof. Let $G=(V,E)$ be a graph and $D_s,D_t \subseteq V$ be two dominating sets with $k \coloneqq \vert D_s\vert = \vert D_t \vert $. 

Define the graph $\widetilde{G}=\left( \widetilde{V}, \widetilde{E} \right)$ with $V_q = \{v_q \mid v\in V\}$ for $q \in [9]$, $a,b \notin V$ and
\begin{equation*}
    \begin{split}
        \widetilde{V} \coloneqq &\, \{ a,b \} \cup \left( \bigcup_{q=1}^9 V_q\right),\\
        \widetilde{E} \coloneqq & \,\{\{a,b\}\}\cup \{\{v_1,u_3\},\{v_2,u_3\}\mid v,u\in V, u\in N_G[v] \}\\
        &\cup \{\{v_2,b\}, \{v_7,b\}, \{v_4,v_3\},\{v_4,v_5\}, \{v_6,v_5\},\{v_7,v_8\}, \{v_9,v_8\}\mid v\in V\}.    
    \end{split}
\end{equation*}
 This graph is bipartite with the partition $A=\{a\}\cup V_1\cup V_2\cup V_4\cup V_6\cup V_7\cup V_9$ and $B = \{b\}\cup V_3\cup V_5\cup V_8$. This can be easily seen, as for each edge in the specification of $\widetilde{E}$, we first mention the vertex of $A$ and then the vertex of $B$. For a vertex set $D\subseteq V$, we define $A_D = \{v_1\mid v\in D\} \cup V_2\cup \{a\}$. 

\begin{claim}\label{claim:off_all_tj_pspace}
     Let $D\subseteq V$. $D$ is a dominating set of~$G$ \iffl $A_D$ is an \offall of~$\widetilde{G}$.
\end{claim}
\begin{pfclaim}
    In the proof of the claim, we abbreviate $A \coloneqq A_D$. 
    Clearly, $\partial A = \{ b \} \cup V_3$. As $d_A(b)\vert =\vert \{a\} \cup V_2 \vert > \vert V_7\vert = d_{\widetilde{V} \setminus A}(b)$, we need not consider $b$ in the following.

    Assume $D$ is a dominating set. Then for each $u\in V$, $N_G[u] \cap D$ is not empty. Therefore, there exists a $v_1\in N_{\widetilde{G}}(u_3)\cap A\cap V_1$. Hence, $d_{\widetilde{V} \setminus A}(u_3) \leq \vert \{u_4\} \cup (N_{V_1}(u_3) \setminus \{v_1\})\vert = \vert N_G[u]\vert < \vert N_G[u]\vert + 1 = \vert N_{V_2}(u_3)\cup \{v_1\} \vert \leq d_{A}(u_3)$. Thus, $A$ is an \offall. 

    Assume $D$ is not a dominating set. So there exists a $u \in V$ such that $N_G[u] \cap D = \emptyset$. Thus, $N_{\widetilde{G}}(u_3)\cap A\cap V_1$ is also empty. Therefore, $d_{\widetilde{V} \setminus A}(u_3) = \vert \{u_4\} \cup N_{V_1}(u_3)\vert = \vert N_G[u]\vert + 1 > \vert N_G[u]\vert  = \vert N_{V_2}(u_3)\vert = d_{A}(u_3)$. Hence, $A$ is not an \offall. 
\end{pfclaim} 
With the same arguments as in the proofs above, $D_s=D_1,\ldots,D_{\ell}=D_t$ is a \dstjumpseq of~$G$ only if $A_{D_s}=A_{D_1},\ldots, A_{D_{\ell}}=A_{D_t}$ is an \offalltjumpseq of~$\widetilde{G}$. 

Let $A_{D_s}= A_1, \ldots, A_{\ell} = A_{D_t}$ be an \offalltjumpseq of~$\widetilde{G}$. We will show by induction that, for each $i\in [\ell]$, there exists a dominating set $D_i \subseteq V$ such that $A_i = A_{D_i}$ and that $A_i\triangle A_{i-1}=\{v_1\in V_1 \mid v \in D_i\triangle D_{i-1}\}$ if $i>1$. For $A_1$, this is true. So assume that for $A_i$ there exists a $D_i$ such that $A_i=A_{D_i}$. This implies $A_i\cap (\{ b \}\cup (\bigcup_{q=3}^9 V_q))= \emptyset$ and $V_2 \subseteq A_i$. Let $x\in A_i\setminus A_{i+1}$ and $y\in A_{i+1}\setminus A_i$. It is enough to show that $x,y \in V_1$. Clearly, $x \notin \{ b \}\cup (\bigcup_{q=3}^9 V_q)$ and $y\notin V_2 \cup \{ a \}$.  If $y=v_q$ for $v \in V$ and $q \in \{6,9\}$, then $v_{q-1}\in \partial A_{i+1}$ and $d_{A_{i+1}}(v_{q-1}) = 1 = d_{\widetilde{V} \setminus A_{i+1}}(v_{q-1})$. For $y=v_5$ (resp. $y=v_8$) with $v \in V$, $z=v_4\in \partial A_{i+1}$ (resp. $z=v_7\in \partial A_{i+1}$) and $d_{A_{i+1}}(z) =  1  = d_{\widetilde{V} \setminus A_{i+1}}(z)$. If there exists some $v \in V$ and $q\in \{4,7\}$ with $v_q\in A_{i+1}$, then $v_{j+1}\in \partial A_{i+1}$ and $d_{A_{i+1}}(v_{q+1}) =  1  = d_{\widetilde{V} \setminus A_{i+1}}(v_{q+1})$. For a $y\in \{b\} \cup V_3$ and $z\in \partial A_{i+1} \cap (V_4 \cup V_7)$, $d_{A_{i+1}}(z) =  1  = d_{\widetilde{V} \setminus A_{i+1}}(z)$. Therefore, $y\in V_1$. Now assume $x \in \{a\} \cup V_2$. Since $V_7\cap A_{i+1}=\emptyset$, $d_{A_{i+1}}(b) =  \vert V \vert <\vert V_7\vert + 1  = d_{\widetilde{V} \setminus A_{i+1}}(b)$. This would contradict the \offall property of $A_{i+1}$. Therefore, $x,y\in V_1$, and for every $i\in [\ell]$, there exists a $D_i\subseteq V$ such that $A_i=A_{D_i}$.  Because $|D_{i+1}\setminus D_i|=|D_i\setminus D_{i+1}|$, $D_s=D_1,\ldots,D_{\ell}= D_t$ is a \dstjumpseq. Therefore, $(G,D_s,D_t)$ is a \yes-instance of \DSReconf \iffl $(\widetilde{G}, A_{D_s},A_{D_t})$ is a \yes-instance of \offallianceReconf.
\end{pf}

\shortversion{\noindent}
For the remainder of this section, we will consider special versions of \offalls.

\begin{theorem}\label{thm:g_off_all_TJ_PSpace}
\longversion{The problem}           \textsc{G-}\offallianceReconf-TJ is \pspace-complete, even on bipartite graphs.  
\end{theorem}
\begin{figure}[bt]
    \centering
    	
	\begin{tikzpicture}[transform shape]
		      \tikzset{every node/.style={ fill = white,circle,minimum size=0.3cm}}
            \node  at (1.5,0.5)[label={above:$\{u_3\mid u\in N[v]\}$}]{};
            \node [label={above:$=$}] at (1.5,1){};
            \node [label={above:$\{u_1\mid u\in N[v]\}$}] at (-1.5,0.5) {};
            \node [label={above:$=$}] at (-1.5,1){};
            \node [label={above:$V_4 \cup V_5$}] at (-3.5,1.1) {};
            \node [label={above:$=$}] at (-3.5,1){};

			\node[draw,label={above:$v_2$}] (v2) at (0,0) {};
			\node[draw] (v11) at (-1.5,0.75) {};
			\node[draw] (v12) at (-1.5,-0.75) {};
			\node[draw] (v31) at (1.5,0.75) {};
			\node[draw] (v32) at (1.5,-0.75) {};

			\node[draw] (v4) at (-3.5,0.75) {};
			\node[draw] (v5) at (-3.5,-0.75) {};
			\node[draw,label={left:$a_1$}] (a1) at (-5,0.75) {};
			\node[draw,label={left:$a_{2k'+1}$}] (a2) at (-5,-0.75) {};
			\node[draw,label={right:$b_1$}] (b1) at (3,0.75) {};
			\node[draw,label={right:$b_{2k'+1}$}] (b2) at (3,-0.75) {};
   
            \node at (1.5,0.15) {\vdots};
            \node at (-1.5,0.15) {\vdots};
            \node at (-1.5,0.15) {\vdots};
            \node at (-3.5,0.15) {\vdots};
            \node at (-5,0.15) {\vdots};
            \draw (1.5,0) ellipse (15pt and 35pt);
            \draw (-3.5,0) ellipse (15pt and 35pt);
            \draw (-1.5,0) ellipse (15pt and 35pt);			
            \path (v11) edge[-] (v2);
			\path (v12) edge[-] (v2);		
            \path (v31) edge[-] (v2);
			\path (v32) edge[-] (v2);
			\path (v11) edge[-] (v4);
			\path (v12) edge[-] (v4);
			\path (v11) edge[-] (v5);
			\path (v12) edge[-] (v5);
			\path (a1) edge[-] (v4);
			\path (a2) edge[-] (v4);
			\path (a1) edge[-] (v5);
			\path (a2) edge[-] (v5);
			\path (b1) edge[-] (v31);
			\path (b2) edge[-] (v31);
			\path (b1) edge[-] (v32);
			\path (b2) edge[-] (v32);
        \end{tikzpicture}

    \label{fig:g_off_all_TJ_PSpace}
    \caption{Construction for \autoref{thm:g_off_all_TJ_PSpace}, for each $v\in V$.}
\end{figure}
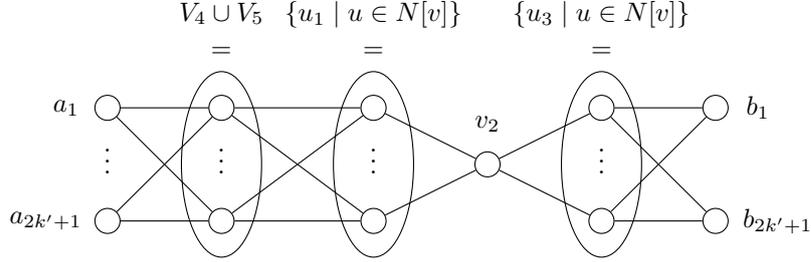
\begin{pf}
For the hardness result, we will again use \DSReconf-TJ. Let $G=(V,E)$ be a graph without isolates and $D_s,D_t$ be dominating sets of~$G$ with $k \coloneqq \vert D_s\vert = \vert D_t \vert$. Then define $V_q \coloneqq \{v_q\mid v\in V\}$ for $q\in [5]$, $k' \coloneqq k + 3 \cdot \vert V\vert $, $V_z\coloneqq\{z_1,\ldots,z_{2k'+1}\}$ for $z\in \{a,b\}$, and $\widetilde{G}=\left(\widetilde{V}, \widetilde{E}\right)$ with
\begin{equation*}
    \begin{split}
        \widetilde{V} \coloneqq{} & V_a\cup V_b \cup \left( \bigcup_{i=1}^5 V_i\right),\\
        \widetilde{E} \coloneqq{} & \{\{v_1 ,u_2\}, \{v_3 ,u_2\} \mid v,u\in V, v\in N[u]\} \cup \{ \{v_1,u_4\}, \{v_1,u_5\} \mid v,u\in V \} \cup{} \\
        & \{ \{ a_j, v_4\}, \{ a_j, v_5\},\{v_3,b_j\} \mid v\in V, j\in [2k' + 1] \}.
    \end{split}
\end{equation*}
For $D \subseteq V$, let $A_D \coloneqq \{v_1\mid v\in D\}\cup V_3 \cup V_4 \cup V_5$. $\widetilde{G}$ is bipartite with the two classes $V_1 \cup V_3 \cup V_a$ and $V_2\cup V_4 \cup V_b$. As before, we describe the edges by first showing the vertex of the first class and then that of the second class, making the bipartiteness evident.

\begin{claim}\label{claim:dom_to_global_off}
    Let $D \subseteq V$. $A_D$ is a global \offall of $\widetilde{G}$ \iffl $D$ is a dominating set of~$G$. 
\end{claim}
\begin{pfclaim}
    To simplify the notation, let $A \coloneqq A_D$. Let $x \in \partial A$. For $x\in V_a\cup V_b$, $d_A(x) \geq \vert V \vert  >0=d_{\widetilde{V} \setminus A}(x)$. If $x=v_1$ with $v\in V$, then $d_A(x) \geq \vert V_4\cup V_5\vert= 2 \cdot \vert V \vert > d_G(v) + 1 \geq d_{\widetilde{V} \setminus A}(x).$

    Let $D$ be a dominating set of~$G$. Thus, for each $v\in V$, $N[v] \cap D\neq \emptyset$. Hence, $d_A(v_2)\geq d_G(v) + 2 > d_G(v) \geq d_{\widetilde{V} \setminus A}(v_2)$. As $\partial A\subseteq V_1 \cup V_2\cup V_a\cup V_b$, $A$ is an \offall. 

    Assume $D$ is not a dominating set. Then there exists a $v\in V$ such that $N[v]\cap D = \emptyset$. Thus, $d_A(v_2)= d_G(v)+1= d_{\widetilde{V} \setminus A}(v_2)$. Since $v_2 \in \partial A$, $A$ is not an \offall. 
\end{pfclaim}

Therefore, if there exists a \dstjumpseq $D_s = D_1,\ldots, D_\ell = D_t$, there exists a global \offalltjumpseq $A_{D_s}=A_{D_1},\ldots,A_{D_\ell}=A_{D_t}$ of~$\widetilde{G}$. 

Let $A_{D_s}=A_1,\ldots, A_\ell = A_{D_t}$ be a global \offalltjumpseq of~$\widetilde{G}$. Since each vertex in $V_3,V_4,V_5$ has degree at least $ 2k'+1$, these vertices must be in each global \offall with at most $k'$ vertices.

\begin{claim}\label{claim:g_off_all_stepsV1}
    There exists a global \offalltjumpseq $A_{D_s}=A'_1,\ldots, A'_{p} = A_{D_t}$ with $p < \ell$ and $A'_i \cap (V_a\cup V_b \cup V_2 ) = \emptyset$ for each $i \in [p]$. 
\end{claim}
\begin{pfclaim}
    Assume there exists a $u\in V_a\cup V_b$ such that there exists a $q\in [\ell]$ with $u\in A_q$. Let $A_i,\ldots,A_j$ (with $i + 2 \leq j$) be the shortest subsequence of consecutive sequence members of $A_{1},\ldots,A_\ell$ such that there exists a $u \in \bigcap_{z=i+1}^{j-1} A_z \neq \emptyset $ and $u\notin A_i \cup A_j$. We will now show that we can substitute or delete parts of the subsequence such that we decrease the number of $r\in [\ell]$ with $u \in A_q$.
    
    Let $v \in A_i\setminus A_{i+1}$, $x\in A_{i+1}\setminus A_{i+2}$ and $y\in A_{i+2} \setminus A_{i+1}$. If $v=y$ or $u=x$, then $A_i,A_{i+2},\ldots,A_j$ is a shorter global \offalltjumpseq. If both cases do not hold, define $A'_{i+1} \coloneqq \left(A_i \cup \{y\}\right) \setminus \{x\}$. Clearly, $A_i,A'_{i+1},A_{i+2},\ldots,A_j$ is a global \offalltjumpseq if $A'_{i+1}$ is a global \offall. Since $V_3 \cup V_4 \cup V_5 \subseteq \left(\bigcap_{t=i}^j A_t\right)$, $u,v,x,y\notin V_3 \cup V_4 \cup V_5 $ and $A'_{i+1}$ is a dominating set. Further, for $w\in V_a\cup V_b$, $d_{A'_{i+1}}(w) \geq \vert V \vert > 0 = d_{\widetilde{V} \setminus A'_{i+1}}(w)$.  If $w\in \left( \partial A'_{i+1}\right) \setminus N_{\widetilde{G}}[x]$, then $d_{A'_{i+1}}(w) \geq d_{A_{i}}(w) > d_{\widetilde{V} \setminus A_{i}}(w) \geq d_{\widetilde{V} \setminus A'_{i+1}}(w)$. As $A_{i+2} =  \left( A'_{i+1}\cup \{u\}\right) \setminus \{v\}$ and $N(u)\subseteq V_3\cup V_4 \cup V_5 \subseteq A'_{i+1}$, $d_{A'_{i+1}}(w) \geq d_{A_{i+2}}(w) >d_{\widetilde{V} \setminus A_{i+2}}(w) \geq d_{\widetilde{V} \setminus A'_{i+1}}(w)$ for $w\in N[x]$. Thus, $A'_{i+1}$ is an \offall. 
    From now on, we can assume $A_r \cap (V_a\cup V_b) = \emptyset$ for $r\in [\ell]$. 
    
    Assume there exists a $u\in V$ and a $r\in [\ell]$ with $u_2\in A_r$. Let $A_i,\ldots,A_j$ (with $i + 2 \leq j$) be the shortest subsequence of consecutive sequence members of $A_{1},\ldots,A_\ell$, such that $u_2 \in \bigcap_{r=i+1}^{j-1} A_r \neq \emptyset $ and $u_2\notin A_i \cup A_j$. As before, we try to decrease the length of this subsequence. Let $v \in A_i\setminus A_{i+1}$, $x\in A_{i+1}\setminus A_{i+2}$ and $y\in A_{i+2} \setminus A_{i+1}$. As above, if $v=y$ or $u=x$, then $A_i,A_{i+2},\ldots,A_j$ is a shorter global \offalltjumpseq. Assume there exists a $w\in N_G[u]$ such that $x=w_1$. Then define $A'_{i+1} \coloneqq \left( A_i\cup \{ y \}\right)\setminus \{v\}$. Since $V_3 \cup V_4 \cup V_5 \subseteq \left(\bigcap_{t=i}^j A_t\right)$, $u,v,x,y\notin V_3 \cup V_4 \cup V_5 $ and $A'_{i+1}$ is a dominating set. Since $A_{i+2}=\left( A'_{i+1}\cup \{u_2\}\right)\setminus \{ x\}$, it is enough to show that $A'_{i+1}$ is an \offall to prove that $A_i, A'_{i+1}, A_{i+2},\ldots, A_j$ is a global \offalltjumpseq. As $V_3 \cup V_4 \cup V_5\subseteq A'_{i+1}$, $d_{A'_{i+1}}(z)>d_{\widetilde{V} \setminus A'_{i+1}}(z)$ holds for $z \in V_1 \cup V_a\cup V_b$. For $z \in V_2 \setminus N[v]$, $d_{A'_{i+1}}(z) \geq d_{A_i}(z) > d_{\widetilde{V} \setminus A_i}(z)= d_{\widetilde{V}\setminus A'_{i+1}}(z)$. Since $u_2 \in A_{i+2}$ and $N(u_2)\subseteq V_1 \cup V_3$,  for $z\in V_2\cap N[v]$, $d_{A'_{i+1}}(z) \geq d_{A_{i+2}}(z) > d_{\widetilde{V} \setminus A_{i+2}}(z)= d_{\widetilde{V}\setminus A'_{i+1}}(z)$. Hence, $A'_{i+1}$ is a global \offall and $A_i,A'_{i+1},A_{i+2},\ldots,A_j$ is a global \offalltjumpseq.

    Analogously to the existence of a $u\in V_a\cup V_b$, $A_i,A'_{i+1}\coloneqq \left(A_i \cup \{y\}\right) \setminus \{x\},A_{i+2},\ldots,A_j$ is a global \offalltjumpseq.
    \end{pfclaim}
    The claim provides that there \longversion{exists}\shortversion{is} a global \offalltjumpseq $A_{D_s} = A_1, \ldots, A_{\ell} = A_{D_t}$ such that for each $t\in [\ell]$, there exists a $D_i\subseteq V$ with $A_i=\{v_1\mid v\in D_i\} \cup V_3 \cup V_4 \cup V_5= A_{D_i}$. By \autoref{claim:dom_to_global_off}, $D_i$ is a dominating set and $D_s=D_1,\ldots,D_\ell=D_t$ is a \dstjumpseq. 
\end{pf}    

The next theorem not only considers a different graph class (compared to the previous theorem), but also supplements the preceding subsection.

\begin{theorem}\label{thm:g_off_all_TS_Pspace}
\longversion{The problems}     \textsc{G-}\offallianceReconf-TS  and \textsc{G-}\offallianceReconf-TJ are \pspace-complete, even on chordal graphs. 
\end{theorem}

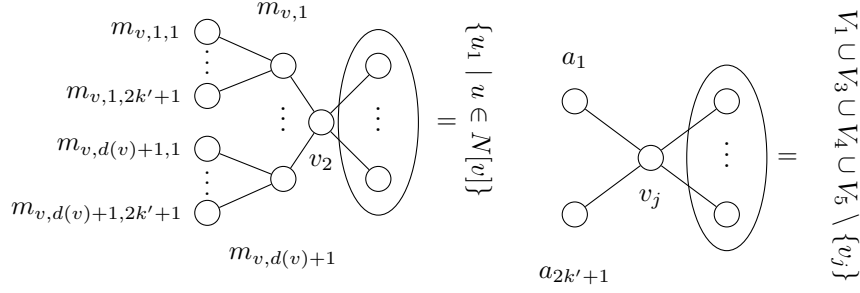
\begin{figure}[bt]
    \centering
    	
\begin{subfigure}[t]{.51\textwidth}
    \centering
    	
	\begin{tikzpicture}[transform shape]
		      \tikzset{every node/.style={ fill = white,circle,minimum size=0.3cm}}
            \node  at (0.25,0)[label={[rotate=-90]above:$\{u_1\mid u\in N[v]\}$}]{};
            \node [label={above:$=$}] at (1.15,-0.5){};
            \node at (-2,0.9) {\small\vdots};
            \node at (-2,-0.7) {\small\vdots};
            \node at (0.25,0.15) {\vdots};
            \node at (-1,0.15) {\vdots};

			\node[draw,label={below:$v_2$}] (v2) at (-0.5,0) {};
			\node[draw,label={above:$m_{v,1}$}] (m1) at (-1,0.75) {};
			\node[draw,label={below:$m_{v,d(v)+1}$}] (m2) at (-1,-0.75) {};
			\node[draw,label={left:$m_{v,d(v)+1,1}$}] (m21) at (-2,-0.35) {};
			\node[draw,label={left:$m_{v,d(v)+1,2k'+1}$}] (m22) at (-2,-1.2) {};
			\node[draw,label={left:$m_{v,1,1}$}] (m11) at (-2,1.2) {};
			\node[draw,label={left:$m_{v,1,2k'+1}$}] (m12) at (-2,0.35) {};
			\node[draw] (v11) at (0.25,0.75) {};
			\node[draw] (v12) at (0.25,-0.75) {};
            \draw (0.25,0) ellipse (15pt and 35pt);			
            \path (m1) edge[-] (v2);
			\path (m2) edge[-] (v2);
			\path (v11) edge[-] (v2);
			\path (v12) edge[-] (v2);			
            \path (m1) edge[-] (m11);
			\path (m1) edge[-] (m12);			
            \path (m2) edge[-] (m21);
			\path (m2) edge[-] (m22);
        \end{tikzpicture}

    \subcaption{Gadget that verifies that each vertex in the original graph is dominated ($v\in V$).}
    \label{fig:g_off_all_domination_gadget}
\end{subfigure}\qquad
\begin{subfigure}[t]{.40\textwidth}
    \centering
    	
	\begin{tikzpicture}[transform shape]
			\tikzset{every node/.style={ fill = white,circle,minimum size=0.3cm}}
            \node  at (0.6,0)[rectangle,label={[rotate=-90]above:$V_1\cup V_3\cup V_4\cup V_5\setminus\{v_j\}$}]{};
            \node [label={above:$=$}] at (1.8,-0.5){};

			\node[draw,label={below:$v_j$}] (vj) at (0,0) {};
			\node[draw] (v11) at (1,0.75) {};
			\node[draw] (v12) at (1,-0.75) {};
			\node[draw,label={above:$a_1$}] (a1) at (-1,0.75) {};
			\node[draw,label={below:$a_{2k' + 1}$}] (a2) at (-1,-0.75) {};
            \node at (1,0.15) {\vdots};
            \draw (1,0) ellipse (15pt and 35pt);			
            \path (a2) edge[-] (vj);		
            \path (a1) edge[-] (vj);
			\path (v11) edge[-] (vj);
			\path (v12) edge[-] (vj);
            
        \end{tikzpicture}

    \subcaption{Construction to ensure that the vertices in~$V_1$ fulfill the \offall property (with $j\in \{3,4,5\}$, $v\in V$)}
    \label{fig:g_off_all_save_V1_gadget}
    \end{subfigure}

    \caption{Construction for \autoref{thm:g_off_all_TS_Pspace} besides the edges of the clique $V_1$.}
\end{figure}
\begin{pf}
For the hardness result, we will again use \DSReconf-TJ. Let $G=(V,E)$ be a graph and $A_s,A_t$ be dominating sets of~$G$ with $k \coloneqq \vert A_s\vert = \vert A_t \vert$. Define $V_q \coloneqq \{v_q\mid v\in V\}$ for $q\in [5]$, $k' \coloneqq k + 4 \cdot \vert V\vert + 2 \cdot \vert E\vert $, $M_v= \{m_{v,j},m_{v,j,p} \mid j \in [d_G(v)+1], p \in [2k'+1]\}$ for $v \in V$ and $\widetilde{G}=\left(\widetilde{V}, \widetilde{E}\right)$~with
\begin{equation*}
    \begin{split}
        \widetilde{V} \coloneqq{} & \{a_1,\ldots,a_{2k'+1}\} \cup \left( \bigcup_{q=1}^5 V_q\right) \cup \left(\bigcup_{v\in V} M_v\right),\\
        \widetilde{E} \coloneqq {}& \binom{V_1 \cup V_3 \cup V_4\cup V_5}{2}\cup \{\{v_1 ,u_2\} \mid v,u\in V, v\in N[u]\} \cup{}\\
        &\{\{v_2,m_{v,j}\}, \{m_{v,j},m_{v,j,p}\}\mid v\in V, j\in [d_G(v)+1], p\in [2k'+1] \}\cup{}\\
        & \{ \{a_p,v_3\}, \{ a_p, v_4\}, \{ a_p, v_5\} \mid v\in V, p\in [2k' + 1] \}.
    \end{split}
\end{equation*}
$\widetilde{G}$ is a chordal graph: First of all, $V_1 \cup V_3\cup V_4 \cup V_5$ is a clique. Therefore, $\{a_1,\ldots,a_{2k'+1}\}$ is a set of simplicial vertices. Further, for $v\in V$, $j\in [d_G(v) + 1]$  $p\in [2k'+1]$, $m_{v,j,p}$ is a leaf. If we delete these vertices from $\widetilde{G}$, then $m_{v,j}$ is a leaf for $v\in V$ and $j\in [d_G(v)+1]$. The vertices in $V_2$ are simplicial in $\widetilde{G}\left[\bigcup_{q=1}^5 V_q\right]$. Since  $V_1 \cup V_3\cup V_4 \cup V_5$ is a clique, $\widetilde{G}$ is chordal.

For $D \subseteq V$, define $A_D \coloneqq \{v_1\mid v\in D\}\cup V_3 \cup V_4 \cup V_5\cup \{m_{v,j} \mid v\in V, j\in [d_G(v) + 1] \}$. 

\begin{claim}\label{claim:g_off_chordal}
    Let $D \subseteq V$. $A_D$ is a global \offall of~$\widetilde{G}$ \iffl $D$ is a dominating set of~$G$. 
\end{claim}
\begin{pfclaim} Let $D \subseteq V$.
    To simplify the notation, let $A \coloneqq A_D$. Let $x \in \partial A$. Note that $V_3\cup\{m_{v,j} \mid v\in V, j\in [d_G(v) + 1] \} \subseteq A$ is a dominating set. If $v\in V$, $j\in[d_G(v)+1], p\in [2k'+1]$ such that $x=m_{v,j,p}$, then $d_{A}(x)=1>0=d_{\widetilde{G}\setminus A}(x)$. For $x\in \{a_1, \ldots ,a_{2k'+1}\}$, $d_A(x)=3\cdot \vert V\vert >0=d_{\widetilde{V} \setminus A}(x)$.  If $x=v_1$ with $v\in V$, then $d_A(x) \geq 3 \cdot \vert V \vert > d_G(v) + 1 + \vert V_1 \setminus \{ v_1 \} \vert \geq d_{\widetilde{V} \setminus A}(x).$

    Let $D$ be a dominating set of~$G$. Thus, for each $v\in V$, $N[v] \cap D\neq \emptyset$. Hence, $d_A(v_2)\geq d_G(v) + 2 > d_G(v) \geq d_{\widetilde{V} \setminus A}(v_2)$. As $\partial A\subseteq V_1 \cup V_2\cup\{a_1, \ldots ,a_{2k'+1}\}\cup \{m_{v,j,p} \mid j \in [d_G(v)+1], p \in [2k'+1]\}$, $A$ is an \offall. 

    Assume $D$ is not a dominating set. Then, there exists a $v\in V$, such that $N[v]\cap D = \emptyset$. Thus, $d_A(v_2)= d_G(v)+1= d_{\widetilde{V} \setminus A}(v_2)$. Since $v_2 \in \partial A$, $A$ is not an \offall. 
\end{pfclaim}
Therefore, if there exists a \dstjumpseq $D_s = D_1,\ldots, D_\ell = D_t$, there exists a global \offalltjumpseq $A_{D_s}=A_{D_1},\ldots,A_{D_\ell}=A_{D_t}$ of~$\widetilde{G}$. Since the \offalls differ only in the vertices in $V_1$ and $V_1$ is a clique, each reconfiguration step is also a token sliding step.

Let $A_{D_s}=A'_1,\ldots, A'_\ell = A_{D_t}$ be a global \offalltjumpseq of~$\widetilde{G}$. Analogously to \autoref{claim:g_off_all_stepsV1}, we can show that there is a global \offalltjumpseq $A_{D_s}=A_1,\ldots, A_p = A_{D_t}$ with $p<\ell$ such that the sets only differ in vertices in $V_1$. Therefore, for each $i\in[p]$, there exists a $D_i \subseteq V$ with $A_i=A_{D_i}$. By \autoref{claim:g_off_chordal}, $D_s=D_1,\ldots,D_{p}=D_t$ is a \dstjumpseq on $G$.
\end{pf}

Motivated by \cite{RodSig2006}, we add the independence condition on top; this study will also stretch into the next subsection.
The domination condition added then gives fewer surprises.

\begin{theorem}\label{thm:i_off_all_TJ_Pspace}
  \longversion{The problem}     \textsc{Idp}-\offallianceReconf-TJ is \pspace-complete, even on bipartite graphs. 
\end{theorem}

\begin{figure}[bt]
    \centering
\begin{subfigure}[t]{.45\textwidth}
    \centering
    	
	\begin{tikzpicture}[transform shape]
		      \tikzset{every node/.style={ fill = white,circle,minimum size=0.3cm}}
            \node  at (2,0)[label={[rectangle,rotate=-90]above:$\{u_3\mid u\in N[v]\}$}]{};
            \node [label={above:$=$}] at (1.8,-0.5){};
            \node  at (-2,0)[label={[rectangle,rotate=90]above:$\{u_1\mid u\in N[v]\}$}]{};
            \node [label={above:$=$}] at (-1.8,-0.5){};

			\node[draw,label={above:$v_2$}] (v2) at (0,0) {};
			\node[draw] (v11) at (-1,0.75) {};
			\node[draw] (v12) at (-1,-0.75) {};
			\node[draw] (v31) at (1,0.75) {};
			\node[draw] (v32) at (1,-0.75) {};
   
            \node at (1,0.15) {\vdots};
            \node at (-1,0.15) {\vdots};
            \draw (1,0) ellipse (15pt and 35pt);
            \draw (-1,0) ellipse (15pt and 35pt);			
            \path (v31) edge[-] (v2);
			\path (v32) edge[-] (v2);
			\path (v11) edge[-] (v2);
			\path (v12) edge[-] (v2);
        \end{tikzpicture}

    \subcaption{Gadget that verifies that each vertex in the original graph is dominated ($v\in V$).}
    \label{fig:i_off_all_TJ_Domination_gadget}
\end{subfigure}\qquad
\begin{subfigure}[t]{.45\textwidth}
    \centering
    	
	\begin{tikzpicture}[transform shape]
			\tikzset{every node/.style={ fill = white,circle,minimum size=0.3cm}}
            
            \node[draw,label={above:$v_{3}$}] (v3) at (0,0) {};
			\node[draw,label={above:$v_5$}] (v5) at (1,0) {};
			\node[draw,label={below:$v_4$}] (v4) at (1,-1) {};
			\node[draw,label={above:$v_{6}$}] (v6) at (2,0) {};
			\node[draw,label={above:$v_{7}$}] (v7) at (3,0) {};
			\node[draw,label={above:$v_{8}$}] (v8) at (4,0) {};		
            \path (v3) edge[-] (v5);
			\path (v5) edge[-] (v4);			
            \path (v5) edge[-] (v6);
			\path (v7) edge[-] (v6);
			\path (v8) edge[-] (v7);
        \end{tikzpicture}

    \subcaption{Gadget that verifies that $v_3$ stays in the \defall ($v\in V$).}
    \label{fig:i_off_all_TJ_v3_stays}
    \end{subfigure}

    \caption{Construction for \autoref{thm:i_off_all_TJ_Pspace}.}
\end{figure}
\begin{pf}
We will again use \DSReconf-TJ. Therefore, let $G=(V,E)$ be a graph and $D_s,D_t\subseteq V$ be dominating sets with $k \coloneqq \vert D_s\vert =\vert D_t\vert$. Define $V_q \coloneqq \{v_q\mid v\in V\}$ for $i\in [8]$ and $\widetilde{G}\coloneqq\left(\widetilde{V},\widetilde{E}\right)$ with
\begin{equation*}
    \begin{split}
        \widetilde{V} \coloneqq{}& \bigcup_{q=1}^8 V_q\\
        \widetilde{E} \coloneqq{}& \{\{v_1,u_2\}, \{v_3,u_2\} \mid v,u\in V, u\in N[v]\} \cup{} \\
        & \{\{v_3,v_5\} \mid v\in V\} \cup \{\{v_q,v_{q+1}\} \mid v\in V, q\in \{4,5,6,7\}\}.
    \end{split}
\end{equation*}
This graph is bipartite with the partition $V_1\cup V_3\cup V_4\cup V_6 \cup V_8$ and $V_2 \cup V_5\cup V_7$.
For a set $D\subseteq V$, we define $A_D\coloneqq\{v_1\mid v\in D\} \cup V_3 \cup V_4$.

\begin{claim}\label{claim:in_off_all_ts_pspace}
     Let $D\subseteq V$. $D$ is a dominating set of~$G$ \iffl $A_D$ is an independent \offall of~$G'$.
\end{claim}
\begin{pfclaim}
     Let $D\subseteq V$. For consistency of notation, let $A\coloneqq A_D$. Clearly, $A$ is an independent set of~$\widetilde{G}$. Thus, we only have to verify that $A$ is an \offall of~$\widetilde{G}$ \iffl $D$ is a dominating set of~$G$. The boundary of $A$ is given by $\partial A= V_2 \cup V_5$. For $v_5 \in V_5$, $d_A(v_5)= \vert \{v_3,v_4\}\vert > \vert \{ v_6\} \vert = d_{\widetilde{V} \setminus A}(v_5)$. This leaves us to check $V_2$.
     
     Assume $D$ is a dominating set. Hence, for each $u\in V$, there exists a $v\in N[u]$. So for each $u_2\in V_2$, there exists a $v_1\in V_1 \cap N_{A}(u_2)$. This implies $d_{A}(u_2) \geq d_{G}(u)+2 > d_{G}(u) \geq d_{ \widetilde{V} \setminus A}(u_2)$ for each $u_2\in V_2$.
     
     If $D$ is not a dominating set, then there exists a $u \in V$ with $D \cap N[u]= \emptyset$. This would imply $d_A(u_2)=d_G(u)+1=d_{\widetilde{V} \setminus A}(u_2)$. Therefore, $A$ is not an \offall if $D$ is not a dominating set.
\end{pfclaim}
This implies that if there is a \dstjumpseq $D_s=D_1,\ldots,D_{\ell}=D_t$ of~$G$, then there is an independent \offalltjumpseq $A_{D_s}= A_{D_1},\ldots, A_{D_{\ell}}=A_{D_t}$ of~$\widetilde{G}$.

Assume that there is an independent \offalltjumpseq $A_{D_s} = A_1,\ldots, A_{\ell}=A_{D_t}$. Let $i\in [\ell]$ be such that there exists a dominating set $D_i\subseteq V$ of~$G$ with $A_i=A_{D_i}$. For $i = 1$ and $i = \ell$, this holds. Let $x\in A_i\setminus A_{i+1}$ and $y\in A_{i+1} \setminus A_i$. This implies $x\notin V_2 \cup V_5 \cup V_6 \cup V_7\cup V_8 $ and $y \notin V_3 \cup V_4$. If there exists a $v\in V$ and $q\in \{5,6,7,8\}$ such that $y=v_q$, then either $v_6$ or $v_7$ will be in $\partial A_{i+1}$ but will not fulfill the \offall property. If $x=v_q\in V_3 \cup V_4$ ($v\in V$), then $v_5\in \partial A_{i+1}$ with $d_{A_{i+1}}(v_5)=1 < 2 = d_{\widetilde{V} \setminus A_{i+1}}(v_5)$. Thus, $x \in V_1$. For $y=v_2\in V_2$, $A_{i+1}$ is no independent set, since $v_2,v_3\in A_{i+1}$ and $\{v_3,v_2\}\in \widetilde{E}$. Therefore, $x,y\in V_1$. This implies that there exists a set $D_{i+1} \subseteq V$ with $A_{i+1}=A_{D_{i+1}}$. By \autoref{claim:in_off_all_ts_pspace}, $D_{i+1}$ is a dominating set. Also, $\vert A_i\triangle A_{i+1}\vert =\vert D_i\triangle D_{i+1}\vert $.
\end{pf}    

\begin{theorem}\label{thm:pow_all_TJ_Pspace}
      \longversion{\textsc{Pow} \AllReconf}\shortversion{\textsc{PA-Reconf}}-TJ is \pspace-complete, even on bipartite graphs. 
\end{theorem}

\begin{figure}[bt]
    \centering
    	
\begin{subfigure}[t]{.43\textwidth}
    \centering
    	
	\begin{tikzpicture}[transform shape, scale=0.9]
		      \tikzset{every node/.style={ fill = white,circle,minimum size=0.3cm}}
            \node  at (2.5,0)[label={[rectangle,rotate=90]above:$\{u_3\mid u\in N(v)\}$}]{};
            \node [label={above:$=$}] at (1.8,-0.5){};
            \node  at (-2,0)[label={[rectangle,rotate=90]above:$\{u_1\mid u\in N[v]\}$}]{};
            \node [label={above:$=$}] at (-1.8,-0.5){};

			\node[draw,label={above:$v_2$}] (v3) at (0,0) {};
			\node[draw] (m1) at (-1,0.75) {};
			\node[draw] (m2) at (-1,-0.75) {};
			\node[draw] (v11) at (1,0.75) {};
			\node[draw] (v12) at (1,-0.75) {};
   
            \node at (1,0.15) {\vdots};
            \node at (-1,0.15) {\vdots};
            \draw (1,0) ellipse (15pt and 35pt);
            \draw (-1,0) ellipse (15pt and 35pt);			
            \path (m1) edge[-] (v3);
			\path (m2) edge[-] (v3);
			\path (v11) edge[-] (v3);
			\path (v12) edge[-] (v3);
        \end{tikzpicture}

    \subcaption{Gadget that verifies that each vertex in the original graph is dominated ($v\in V$).}
    \label{fig:powall_TJ_Space_gadget}
\end{subfigure}\qquad
\begin{subfigure}[t]{.43\textwidth}
    \centering
    	
	\begin{tikzpicture}[transform shape, scale=0.9]
		      \tikzset{every node/.style={ fill = white,circle,minimum size=0.3cm}}
            \node[rectangle] at (-0.4,0) {\footnotesize $v_p$};
            \node[rectangle] at (1,-0.3) {\footnotesize $m_{v,p,1}$};
            \node[rectangle] at (1,1.3) {\footnotesize $m_{v,p,2}$};
            \node[rectangle] at (1.7,0.3) {\footnotesize $m_{v,p,3}$};
            \node[rectangle] at (2.7,1) {\footnotesize $m_{v,p,4}$};
            \node[rectangle] at (4,0) {\footnotesize$ m_{v,p,5}$};
            \node[rectangle] at (2.7,-1) {\footnotesize $m_{v,p,6}$};
            
			\node[draw] (vp) at (0,0) {};
			\node[draw] (m1) at (1,0) {};
			\node[draw] (m2) at (1,1) {};
			\node[draw] (m3) at (2,0) {};
			\node[draw] (m4) at (2.7,0.7) {};
			\node[draw] (m5) at (3.4,0) {};
			\node[draw] (m6) at (2.7,-0.7) {};
            \path (m1) edge[-] (vp);
            \path (m1) edge[-] (m2);
            \path (m1) edge[-] (m3);
            \path (m3) edge[-] (m4);
            \path (m4) edge[-] (m5);
            \path (m5) edge[-] (m6);
            \path (m6) edge[-] (m3);
            \end{tikzpicture}

    \subcaption{Gadget that verifies that the token stays on~$v_3$ ($v\in V$).}
    \label{fig:pow_all_domination_gadget}
\end{subfigure}

    \caption{Construction for \autoref{thm:pow_all_TJ_Pspace}.\label{fig:pow_all_dom}}
\end{figure}
\begin{pf}
As before, we will use \textsc{(T-)DS-Reconf}-TJ for the \pspace-hardness proof. Let $G=(V,E)$ be a graph without isolates and $D_s,D_t \subseteq V$ be two dominating sets with $k \coloneqq \vert D_s\vert = \vert D_t \vert $. 

Define the graph $\widetilde{G}=\left( \widetilde{V}, \widetilde{E} \right)$ with $V_q = \{v_q \mid v\in V\}$ for $q \in [3]$, and $M_{p,j}\coloneqq\{m_{v,p,j} \mid v\in V\}, M_{j}=M_{2,j} \cup M_{3,j}$ for $ j\in [6], p\in \{2,3\}$ and
\begin{equation*}
    \begin{split}
        \widetilde{V} \coloneqq{} &V_1\cup V_2\cup V_3\cup \left(\bigcup_{j=1}^6 M_{2,j} \cup M_{3,j} \right) \,\\
        \widetilde{E} \coloneqq{} & \big\{\{v_1,u_2\},\{v_3,u_2\}\mid \{v,u\}\in E\,\big\} \cup \big\{\{v_1,v_2\}\mid v\in V\big\} \cup \\
        &  \bigcup_{v\in V, p\in \{2,3\}}\big( \{\{v_p, m_{v,p,1}\}, \{m_{v,p,1},m_{v,p,3}\}, \{m_{v,p,6},m_{v,p,3}\}\}\cup\\ & \qquad\qquad\quad \big\{\{m_{v,p,j},m_{v,p,j+1}\}\mid j\in\{1,3,4,5\}\big\}\big)\,.
    \end{split}
\end{equation*}

 This graph is bipartite with the partition $A=V_1\cup V_3\cup M_{2,1} \cup M_{2,4}\cup M_{2,6} \cup M_{3,2} \cup M_{3,3} \cup M_{3,5}$ and $B = V_2\cup M_{3,1} \cup M_{3,4}\cup M_{3,6} \cup M_{2,2} \cup M_{2,3} \cup M_{2,5}$. The reader can verify this by looking at \autoref{fig:pow_all_dom}. For a vertex set $D\subseteq V$, we define $A_D = \{v_1\mid v\in D\} \cup V_2\cup V_3 \cup M_2$.

\begin{claim}\label{claim:pow_all_tj_pspace}
     Let $D\subseteq V$. $D$ is a dominating set of~$G$ \iffl $A_D$ is a \powall of~$\widetilde{G}$.
\end{claim}
\begin{pfclaim}
    In the proof of the claim, we abbreviate $A \coloneqq A_D$. 
    Clearly, $\partial A = \{v_1\mid v\in \widetilde{V} \setminus D\}\cup M_1$. For each $v_1\in V$, $N(v_1)\subseteq V_2 \subseteq A$. Since $d_A(m_{v,p,1}) = 2 = d_{\widetilde{V}\setminus A}(m_{v,1}) + 1$ for all $v\in V$, $A$ is an \offall. Further, $d_A(v_3) + 1 = d_G(v) > 1 = d_{\widetilde{V} \setminus A}(v_3)$ holds for $v\in V$.
    For $m_{v,p,j}\in M_{p,j}$ with $ j\in [6], p\in \{2,3\}$, $d_A(m_{v,p,2}) + 1 = 1 = d_{\widetilde{V} \setminus A}(m_{v,p,2})$.
    Hence, we only need to consider the vertices in $V_2$.

    Let $D$ be a dominating set. Then, for each $u\in V$ there exists a $v\in N_G[u]\cap D$. So $d_{A}(u_2) + 1 \geq d_G(u) + 1  \geq d_{\widetilde{V} \setminus A}(u_2)$ and $A$ is a defensive/powerful alliance. 

    If $D$ is not a dominating set, then there exists a $u \in V$ such that $N_G[u] \cap D$ is empty. Thus, $d_{A}(u_2) + 1 =  d_G(u) < d_G(u) + 2 = d_{\widetilde{V} \setminus A}(u_2)$. Therefore, $A$ is not a  \powall if $D$ is not a  dominating set.
\end{pfclaim} 
With the same arguments as in the  proofs above, $D_s=D_1,\ldots,D_{\ell}=D_t$ is a \dstjumpseq of~$G$ \iffl $A_{D_s}=A_{D_1},\ldots, A_{D_{\ell}}=A_{D_t}$ is a powerful \alltjumpseq of~$\widetilde{G}$. 

Let $A_{D_s}= A_1, \ldots, A_{\ell} = A_{D_t}$ be an  powerful \alltjumpseq of~$\widetilde{G}$. We will show by induction that, for each $i\in [\ell]$, there exists a dominating set $D_i \subseteq V$ such that $A_i = A_{D_i}$ and that $D_i\triangle D_{i-1}=\{ v \mid v_1 \in A_i\triangle A_{i-1}\}$ if $i>~1$. For $A_1$, this is true. So assume that for $A_i$ there exists a $D_i$ such that $A_i=A_{D_i}$.  Let $x\in A_i\setminus A_{i+1}$ and $y\in A_{i+1} \setminus A_i$. This implies  $x\notin M_1\cup M_3 \cup M_4 \cup M_5 \cup M_6$ and $y \notin V_2 \cup V_3 \cup M_2$.
For $y = m_{v,p,j}$ with $v\in V, p\in \{2,3\},j \in \{1,4,6\}$, $d_{\widetilde{V} \setminus A}(m_{v,p,3}) + 1 = 2 > 1 = d_A(m_{v,p,3})$. Furthermore, $d_{\widetilde{V} \setminus A}(m_{v,p,4}) + 1 = 2 > 1 = d_A(m_{v,p,4})$ for $y = m_{v,p,j}$ with $v\in V, p\in \{2,3\},j \in \{3,5\}$. Thus, $y\in V_1$. If $x\in \{v_p, m_{v,p,2}\}$ with $v\in V, p\in \{2,3\}$, then $d_{\widetilde{V} \setminus A}(m_{v,p,1}) + 1 = 2 > 1 = d_A(m_{v,p,1})$. Therefore, $x,y\in V_1$, and for every $i\in [\ell]$, there exists a $D_i\subseteq V$ such that $A_i=A_{D_i}$.  Because $|D_{i+1}\setminus D_i|=|D_i\setminus D_{i+1}|$, $D_s=D_1,\ldots,D_{\ell}= D_t$ is a \dstjumpseq. Thus, $(G,D_s,D_t)$ is a \yes-instance of \DSReconf \iffl $(\widetilde{G}, A_{D_s},A_{D_t})$ is a \yes-instance of \longversion{\textsc{Pow} \AllReconf}\shortversion{\textsc{PA-Reconf}}-TJ.
\end{pf}    

\begin{theorem}\label{thm:g_pow_all_TJ_Pspace}
      \longversion{\textsc{G-Pow} \AllReconf}\shortversion{\textsc{G-PA-Reconf}}-TJ is \pspace-complete, even on bipartite graphs.
\end{theorem}

\begin{figure}[bt]
\centering
    	
\begin{subfigure}[t]{.46\textwidth}
    \centering
    	
	\begin{tikzpicture}[transform shape]
		      \tikzset{every node/.style={ fill = white,circle,minimum size=0.3cm}}
            \node  at (2,0)[rectangle,label={[rectangle,rotate=-90]above:$\{u_3\mid u\in N(v)\}$}]{};
            \node [label={above:$=$}] at (1.8,-0.5){};
            \node  at (-2,0)[rectangle,label={[rectangle,rotate=90]above:$\{u_1\mid u\in N[v]\}$}]{};
            \node [label={above:$=$}] at (-1.8,-0.5){};

			\node[draw,label={below:$v_2$}] (v3) at (0,0) {};
			\node[draw,label={above:$v_4$}] (v4) at (-0.4,1) {};
			\node[draw,label={above:$v_5$}] (v5) at (0.4,1) {};
			\node[draw] (m1) at (-1,0.75) {};
			\node[draw] (m2) at (-1,-0.75) {};
			\node[draw] (v11) at (1,0.75) {};
			\node[draw] (v12) at (1,-0.75) {};
   
            \node at (1,0.15) {\vdots};
            \node at (-1,0.15) {\vdots};
            \draw (1,0) ellipse (15pt and 35pt);
            \draw (-1,0) ellipse (15pt and 35pt);			
            \path (m1) edge[-] (v3);
			\path (m2) edge[-] (v3);
			\path (v11) edge[-] (v3);
			\path (v12) edge[-] (v3);
			\path (v4) edge[-] (v3);
			\path (v5) edge[-] (v3);
        \end{tikzpicture}

    \subcaption{Gadget that verifies that each vertex in the original graph is dominated ($v\in V$).}
    \label{fig:g_powall_TJ_Space_gadget}
\end{subfigure}\qquad\begin{subfigure}[t]{.48\textwidth}
    \centering
    	
     \begin{tikzpicture}[transform shape]
		      \tikzset{every node/.style={ fill = white,circle,minimum size=0.3cm}}
            \node  at (2.1,0){$V_2$};
            \node [label={above:$=$}] at (1.8,-0.5){};

			\node[draw,label={below:$v_3$}] (v3) at (0,0) {};
			\node[draw,label={left:$m_{v,1}$}] (m1) at (-1,0.75) {};
			\node[draw,label={left:$m_{v,d_G(v+1)1}$}] (m2) at (-1,-0.75) {};
			\node[draw] (v11) at (1,0.75) {};
			\node[draw] (v12) at (1,-0.75) {};
            \node at (1,0.15) {\vdots};
            \node at (-1,0.15) {\vdots};
            \draw (1,0) ellipse (15pt and 35pt);			
            \path (m1) edge[-] (v3);
			\path (m2) edge[-] (v3);
			\path (v11) edge[-] (v3);
			\path (v12) edge[-] (v3);	
        \end{tikzpicture}

    \subcaption{Gadget that ensures that the tokens stay in~$V_3$ (with $v\in V$).}
    \label{fig:g_pow_all_V_3_gadget}
\end{subfigure}
\qquad
\begin{subfigure}[t]{.91\textwidth}
    \centering
    	
     \begin{tikzpicture}[transform shape]
		      \tikzset{every node/.style={ fill = white,circle,minimum size=0.3cm}}

            \node at (0,1) {\ldots};
            \node at (-2.5,-1) {\ldots};
            \node at (2,-1) {\ldots};
            \node  at (0,1.4){$V_1$};
            \node at (1,-1.5) {\small$b_{n-k+1}$};
            \node at (-4.6,-2.5) {\small$c_{1}$};
            \node at (-3.4,-2.5) {\small$c_{2}$};
            \node at (-1.65,-2.5) {\small$c_{2(n-k)-1}$};
            \node at (-0.35,-2.5) {\small$c_{2(n-k)}$};
			\node[draw,label={right:\small$a$}] (a) at (0,0) {};
			\node[draw,label={left:\small$b_1$}] (b1) at (-4,-1) {};
			\node[draw,label={right:\small$b_{n-k}$}] (b2) at (-1,-1) {};
			\node[draw] (b3) at (1,-1) {};
			\node[draw,label={below:\small$b_{n}$}] (b4) at (3,-1) {};
			\node[draw] (c21) at (-1.6,-2) {};
			\node[draw] (c22) at (-0.4,-2) {};
			\node[draw] (c11) at (-4.6,-2) {};
			\node[draw] (c12) at (-3.4,-2) {};
			\node[draw] (v11) at (0.75,1) {};
			\node[draw] (v12) at (-0.75,1) {};
            \draw (0,1) ellipse (35pt and 20pt);			
            \path (b1) edge[-] (a);
			\path (b2) edge[-] (a);		
            \path (b3) edge[-] (a);
			\path (b4) edge[-] (a);
			\path (v11) edge[-] (a);
			\path (v12) edge[-] (a);			
            \path (b1) edge[-] (c11);
			\path (b1) edge[-] (c12);			
            \path (b2) edge[-] (c21);
			\path (b2) edge[-] (c22);
        \end{tikzpicture}

    \subcaption{Gadget that verifies that the token stays at $a$.}
    \label{fig:g_pow_all_domination_gadget}
\end{subfigure}

    \caption{Constructions for \autoref{thm:g_pow_all_TJ_Pspace}.\label{fig:g_pow_all}}
\end{figure}

\begin{pf}
We use again \DSReconf-TJ for the \pspace-hardness. Let $G=(V,E)$ be a graph  with $n=\vert V \vert $  without isolates and $D_s,D_t \subseteq V$ dominating sets with $k \coloneqq \vert D_s \vert = \vert D_t \vert$. We assume $k\notin\{0, n-1, n\}$ (for $k=n-1$: If both sets a dominating sets we put the token in $D_s \setminus D_t$ to the vertex in $D_t \setminus D_s$). Otherwise, it is a trivial instance. Define $\widetilde{G}=(\widetilde{V},\widetilde{E})$ with $V_q\coloneqq \{v_q \mid v\in V\}$ for $q\in [5]$, $M_v= \{m_{v,j}\mid j\in [d_G(v)+2]\}$ and 
    \begin{equation*}
        \begin{split}
            \widetilde{V} \coloneqq{}& \left( \bigcup_{q=1}^5 V_q \right)\cup \left( \bigcup_{v\in V} M_v\right) \cup \{a\} \cup \{b_j \mid j\in[n]\} \cup \{c_j \mid j\in[2(n - k)]\},\\
            \widetilde{E} \coloneqq{}& \{\{v_1, a\}, \{v_4,v_2\}\mid v\in V \} \cup \{\{v_1,u_2\},\{v_3,u_2\}\mid v,u\in V,\, u \in  N_G[v]\} \cup{}\\
             & \{\{v_3,m_{v,j}\}\mid v \in V, j\in [d_G(v)+2]\}\cup\{\{b_j,a\}\mid j\in [n] \} \cup{} \\
             &\{\{b_j,c_{2j-1}\},\{b_j,c_{2j}\}\mid j\in [n-k] \}.
        \end{split}
    \end{equation*}
The reader is invited to check out \autoref{fig:g_pow_all} for the construction. 
The constructed graph~$\widetilde{G}$ is bipartite as $B_1\coloneqq V_1 \cup V_3 \cup V_4 \cup V_5 \cup \{b_j \mid j\in[n]\}$ and $B_2 \coloneqq V_2 \cup \left( \bigcup_{v\in V} M_v\right) \cup \{a\} \cup \{c_j \mid j \in [2(n - k)]\}$ form the partition classes. This can be easily checked, since for each edge in the definition of $\widetilde{E}$, we first mention the vertex of $B_1$ and then the vertex of $B_2$. Furthermore, define $A_D \coloneqq  \{v_1\mid v\in D\} \cup V_2\cup V_3 \cup \{a\} \cup \{b_j \mid j \in [n-k]\} $. Hence, $\vert A_{D_s} \vert = \vert A_{D_t} \vert = k + 2\cdot n +n-k+1=3n+1$.

\begin{claim}\label{claim:g_pow_all_tj_pspace}
        Let $D\subseteq V$. $D$ is a dominating set of~$G$ \iffl $A_D$ is a global \powall of~$\widetilde{G}$.
    \end{claim}
    \begin{pfclaim}
        For the proof of this claim, we abbreviate $A\coloneqq A_D$. Since $V \setminus A \subseteq V_4 \cup V_5 \cup \left( \bigcup_{v\in V} M_v\right) \cup \{c_{2j-1},c_{2j} \mid j \in [n-k]\} \cup \{ b_j\mid j\in [n]\setminus [n-k]\}\subseteq N(V_3 \cup \{a\} \cup \{b_j \mid j \in [n-k]\})$, $A$ is a dominating set.
        
        As $N_{\widetilde{G}}(V_1\cup V_4 \cup V_5)=V_2 \cup \{a\}$, we do not have to check $V_1, V_4$ and $V_5$ for the \powall property. Further, $N_{\widetilde{G}}(c_{2j-1}) = \{b_j\} =N_{\widetilde{G}}(c_{2j})\subseteq A$ for each $j\in [n-k]$. For $j\in [n]\setminus [n-k]$, $N_{\widetilde{G}}(b_j)=\{a\} \subseteq A$. Since also $N(M_v)=\{ v_3 \}$ for all $v\in V$, $A$ is an \offall.

        For $a\in A$, $d_{A}(a) + 1=k+(n-k)+1 > n = d_{\widetilde{V} \setminus A}$. The fact $d_{A}(v_3) + 1 = d_G(v) + 2 = d_{\widetilde{V} \setminus A}(v_3)$ for all $v\in V$ implies that we only need to consider $V_2$ for $A$ to be a \powall. 

        Let $D$ be a dominating set. Then for each $v\in V$, there exists a $u\in N_G[v]\cap D$. Hence, $d_{A}(v_2) + 1 \geq d_G(v) + 2 \geq d_{\widetilde{V} \setminus A}(v_2)$.
        Thus, $A$ is a global \powall.

        Assume $D$ is not a dominating set. So, there exists a $v\in V$ such that $N_G[v]\cap D= \emptyset$. Therefore, $d_{A}(v_2) + 1 = d_G(v) + 1 < d_G(v) + 3 =  d_{\widetilde{V} \setminus A}(v_2)$ implies that $A$ is not a \powall.    
\end{pfclaim}
With the same arguments as in the previous proofs, $D_s=D_1,\ldots,D_{\ell}=D_t$ is a \dstjumpseq on $G$ \iffl $A_{D_s}=A_{D_1},\ldots, A_{D_{\ell}}=A_{D_t}$ is an global powerful \alltjumpseq on~$\widetilde{G}$. 

        Let $A_{D_s}=A_1,\ldots,A_{\ell}= A_{D_t}$ be a global powerful \alltjumpseq on~$\widetilde{G}$. 

        \begin{claim}\label{cla:glo_pow_all_tj_V2}
            There exists a \alltjumpseq $A_{D_s}=A'_1,\ldots,A'_{\ell'}= A_{D_t}$ with $\ell' < \ell$ such that, for all $i\in [\ell']$, $\{a\} \cup V_2 \cup V_3 \cup \{b_j \mid j \in [n-k]\} \subseteq A'_i$.
        \end{claim}
        \begin{pfclaim}
            We will only consider $V_2$. The other cases can be treated analogously. Assume there exists a $v \in V$ and an $i\in [\ell]$ such that $v_2 \notin A_i$. Observe that, for each $i\in [\ell]$ with $v_2 \notin A_i$, $v_4,v_5 \in A_i$; otherwise, the \defall would not be global. Then define for all $i\in [\ell]$, $A'_i \coloneqq (A_i \setminus \{v_5\}) \cup \{v_2\}$ if $v_2\notin A_i$ and $A_i'\coloneqq A_i$, otherwise. Now we want show that, for each $t\in [\ell-1]$, $A'_i$ is a powerful global alliance and either $A'_i$ can be transformed into $A'_i$ by a token jumping step or $A'_i= A_i$. 

            Let $i\in [\ell]$. For $v_2\in A_i$, $A'_i$ is clearly a  global \powall. So assume $v_2\notin A_i$. Since $N(v_5) =\{v_2\}$, $d_{A'_i}(x)\geq d_{A_i}(x)$ and $d_{\widetilde{V} \setminus A_i}(x) \geq d_{\widetilde{V} \setminus A'_i}(x)$ for $x \in \widetilde{V} \setminus \{v_2,v_5\}$. Therefore, we only need to consider the \powall property for $v_2,v_5$. As $v_2\in \partial A_i$ and $A_i$ is an \offall,
            $d_{A'_i}(v_2) + 1 = d_{A_i}(v_2) \geq d_{\widetilde{V} \setminus A_i}(v_2) +1 = d_{\widetilde{V} \setminus A'_i}(v_2)$. Since $A_i$ is a \defall and $v_5 \in A_i$, $d_{A'_i}(v_5) = d_{A_i}(v_5) + 1 \geq d_{\widetilde{V} \setminus A_i}(v_5) = d_{\widetilde{V} \setminus A'_i}(v_5) + 1$. Hence, $A'_i$ is a \powall. Furthermore, $A'_i$ is global, as $N[v_5]\subseteq N[v_2]$.

            Let $x\in A_i\setminus A_{i+1}$ and $y \in A_{i+1} \setminus A_i$. If $\{v_2,v_5\} \cap \{ x,y\} = \emptyset $, $A'_{i}$ can be transformed into  $A'_{i+1}$ by a token jumping move, trivially. If $\{v_2,v_5\} = \{ x,y\}$, $A'_i=A'_{i+1}$. So assume $\vert \{v_2,v_5\} \cap \{ x,y\}\vert = 1 $. Without loss of generality, assume $x\in \{ v_2,v_5\}$. Otherwise swap $A_i$ and $A'_{i}$ with $A_{i+1}$ and $A'_{i+1}$. If $x = v_5$, we can make the same token jumping step as for $A_i$ to $A_{i+1}$ also for $A'_i$ to $A'_{i+1}$. For $x = v_2$, $v_5\in A_i \cap A_{i+1}=A'_i \cap A_{i+1}$, because of being global.  Therefore, $A'_i$ can be transformed to $A'_{i+1}$ as the token jumps from $v_5$ to $y\notin \{v_2,v_5\}$. 
            
            Thus, there exists a global powerful \alltjumpseq which fulfills the properties.    
        \end{pfclaim}

        From now on we assume $A_{D_s}=A_1,\ldots,A_{\ell}= A_{D_t}$ fulfills the properties of \autoref{cla:glo_pow_all_tj_V2}. Since $\vert A_i \setminus (\{a\} \cup V_2 \cup V_3) \vert = n $, $N_{\widetilde{G}}(a) \cap (V_2 \cup V_3) = \emptyset$ and $d_{\widetilde{G}}(a)=2n$, $A_i \setminus (\{a\} \cup V_2 \cup V_3)\subseteq N_{\widetilde{G}}(a)=V_1\cup \{b_i \mid i\in [n]\}$.
        
        We have shown that we can assume that $x,y\in V_1 \cup \{b_j\mid j \in [n-k]\}$. Assume there exists a $j\in [n]\setminus [n-k]$ and an $i\in [\ell] $ with $b_j\in A_i$. We will show that we can transform $A_1 , \ldots , A_{\ell}$  (without extending it) into a global powerful \alltjumpseq such that $b_j$ is not in any of the sets of the sequence.

Therefore, let $p,q\in [\ell]$ be maximal with respect to $q-p$ such that $b_j\notin A_{p-1}\cup A_{q+1}$ but $b_j\in \bigcap_{i=p}^q A_i$. Such a pair $p,q$ exists by our assumption and the fact that $b_j\notin A_s\cup A_t=A_1\cup A_{\ell}$. Let $x\in A_{p-1}\setminus A_p$, $y \in A_p\setminus A_{p+1} $ and $z \in A_{p+1}\setminus A_p$. If $x=z$ and $y=b_j$, then we can delete $A_p,A_{p+1}$ as $A_{p-1}=A_{p+1}$. For $y=b_j$ and $x\neq z$, just delete $A_p$ as the token can directly jump from $x$ to $z$. If $x=z$ and $y\neq b_j$, then just delete $A_p$ since the token can directly jump from $y$ to $b_j$. Now we can assume that $b_j,x ,y,z \in V_1\cup \{b_r\mid r\in [n]\setminus [n-k]\}$ are all different vertices. Define $A'_p \coloneqq (A_p \setminus \{b_j\}) \cup \{z\}$. Then, $d_{A'_p}(a) + 1 = d_{A_p}(a) + 1  = n+1 > n = d_{\widetilde{V} \setminus A_p}(a) = d_{\widetilde{V} \setminus A'_p}(a)$. Further $d_{A'_p}(b_j) = 1 = d_{\widetilde{V} \setminus A'_p}(b_j)$ and $d_{A'_p}(z)+1 > d_{A_p}(z) \leq d_{\widetilde{V} \setminus A_p}(z) + 1 > d_{\widetilde{V} \setminus A_p}(z)$. For the remaining vertices $w\in \widetilde{V} \setminus \{a,b_j,z\}$, $d_{A'_p}(w)\geq d_{A_p}(w)$ as well as $d_{\widetilde{V} \setminus A'_p}(w)\geq d_{\widetilde{V} \setminus A_p}(w)$.  Therefore, $A_1,\ldots A_{p-1}, A'_p, A_{p+1},\ldots, A_{\ell}$ is a global powerful \alltjumpseq and there are less $i\in [\ell]$ with $b_j\in A_i$.

If we perform this procedure inductively, we get a global powerful \alltjumpseq $A_s=A_1,\ldots,A_{\ell'}=A_t$, with only tokens from $V_1$ jumping to $V_1$. Therefore, for each $i \in [\ell']$, there exists a $D_i$ such that $A_i=A_{D_i}$. Furthermore, $\vert D_i \triangle D_{i+1}\vert = 2$ for each $i \in [\ell'-1]$. By \autoref{claim:g_pow_all_tj_pspace}, $D_s=D_1,\ldots,D_{\ell'}=D_t$ is a \dstjumpseq of~$G$.     
\end{pf}

\end{toappendix}

\subsection{\logspace\ Membership Results}

Interestingly, some variants of alliance reconfiguration problems are distinctively easier than \pspace. To prove this, showing membership in \logspace\ suffices, as $\logspace\subsetneq\pspace$ is known by the space hierarchy theorem. We start with a simple combinatorial observation.

\begin{lemma}\label{lem:i_off_ts_iso_edge}
    Let $G=(V,E)$ be a graph and $A,B \subseteq V$ be independent \offalls such that $A$ can be transformed by one token sliding step into~$B$. For $x\in A\setminus B$ and $y\in B\setminus A$, $\{x,y\}\in E$ is an isolated edge.
\end{lemma}

\begin{pf}
     Since  $x\in A\setminus B$ and $y\in B\setminus A$ describe a token sliding step, $\{x,y\} \in E$. As $A$ and~$B$ are independent\longversion{ sets}, $N_B(x) \setminus \{y\}=\emptyset = N_A(x) $ and $N_A(y) \setminus \{x\}=\emptyset = N_B(y)$.  If $d_G(x) > 1$ or $d_G(y) > 1$, this contradicts $B$ and $A$ being \offalls. \longversion{Thus, $N_G[x] = \{x,y\} = N_G[y]$.}
\end{pf}
\noindent
This lemma has one immediate algorithmic consequence.

\begin{proposition}\shortversion{$(*)$}\label{prp:ind_off_ts_logspace}
      \textsc{Idp}-\offallianceReconf-TS${}\in{}$\logspace.  
\end{proposition}
\begin{toappendix}
\begin{pf} \shortversion{[of  \autoref{prp:ind_off_ts_logspace}]}
Let $G=(V,E)$ be graph, $A_s$ be the start configuration and $A_t$ the target configuration. \autoref{lem:i_off_ts_iso_edge} implies that for each $x\in A_s\setminus A_t$ there exists a $y\in A_t \setminus A_s$ such that $\{ x,y\}$ is an isolated edge. Otherwise, this is a trivial \no-instance. For the timed version, this leaves to check if $\vert A_s \setminus A_t \vert < T$ holds and for the other version, we can return the answer immediately.
\end{pf}
\end{toappendix}

\noindent
With little more effort, one can also show the next algorithmic result.

\begin{lemma}\shortversion{$(*)$}
\label{lem:global-ind_off_tj_logspace} For $Y\in\{\textrm{TJ},\textrm{TS}\,\}$,
      \textsc{G-Idp}-\offallianceReconf-$Y\in\logspace$.  
\end{lemma}

\begin{toappendix}
\begin{pf}\shortversion{[of \autoref{lem:global-ind_off_tj_logspace}]}
Let $G=(V,E)$ be a graph and let $A_s, A_t \subseteq V$ be global independent \offalls with $k \coloneqq \vert A_s\vert =\vert A_t\vert$. Let us again take a look at one step of the reconfiguration on~$G$. To this end, let $A,B$ be two global independent \offalls such that $A$ can be transformed to $B$ by one token jumping step and $v\in A\setminus B$ and $u\in B\setminus A$.  Therefore, $u\in V\setminus A = \partial A$ and there exists a $w\in A$ such that $\{u,w\}\in E$. If $w \neq v$, then this would be contradiction to the independence of $B$ ($w,u\in B$). Hence, $u$ is a leaf, as otherwise, $A$ would not be an \offall.

Assume $u\in N(v)$. So $v\in \partial B$. Since $B$ is an \offall, $v$ is either a leaf or there exists an $x\in \left(N(v)\setminus \{u\}\right) \cap B$. The existence of such an $x$ would contradict the independence of~$A$. Therefore, $v$ is a leaf. So, $\{v,u\}\in E$ is an isolated edge and we can use the same argument as in \autoref{prp:ind_off_ts_logspace}. Since each step is a token sliding step, this algorithm also works for the TS version of this problem.
\end{pf}
\end{toappendix}

\subsection{Token Removal and Addition}

Now, we will consider TAR reconfiguration steps. Again, we will derive a number of \pspace-completeness results, but this time, we will provide tight combinatorial links between TAR and TJ to be able to profit from earlier findings. The following proofs are adaptions from Lemma~3 of Bonamy  et al.~\cite{BonDorOuv2021}, but 
we will treat it more abstractly, based on a novel notion that we introduce now. Let $X$ be a set property. We call $X$  \emph{reconfiguration monotone increasing}, or \emph{rmi} for short (resp. \emph{decreasing}, or \emph{rmd} for short), if for each $X$-token jumping step $A,B$ (formally, an $X$-$\text{TJ}$ sequence $A,B$) with $v\in B \setminus A$ (resp. $v\in A \setminus B$), also $A \cup \{v\}$ (resp. $A \setminus \{v\}$) fulfills\longversion{ the property}~$X$.
\longversion{Clearly, m}\shortversion{M}onotone increasing properties as domination are \rmi.

\begin{proposition}\label{pro:ReconfMonoCup}
    Let $X,Y$ be \rmi (resp. \longversion{decreasing}\shortversion{\rmd}) properties. Then the property that $X$ and $Y$ hold is also \rmi (resp. \longversion{decreasing}\shortversion{\rmd}).
\end{proposition}


\begin{theorem}\label{thm:ReconfMonoTAR}
    Let $G=(V,E)$ be a graph, $X$ be a \rmi property on vertex sets and $A_s,A_t \subseteq V$ such that $\vert A_s\vert = \vert A_t \vert = k$ and $A_s,A_t$ have the property~$X$. Then, there exists an $X$-$\text{TJ}$ sequence of length at most $\ell$ \iffl there is an $X$-$\text{TAR}$ with threshold $k+1$ of length at most $2\ell$.
\end{theorem}
\begin{pf}
    Let $A_s=A_1,\ldots, A_{\ell}= A_t$ be an $X$-$\text{TJ}$ sequence in~$G$. Define \shortversion{$v_i\in A_{i+1} \setminus A_i$ for $i\in [\ell - 1]$}\longversion{$v_i$ for $i\in [\ell - 1]$ as the vertex in $A_{i+1} \setminus A_i$}. As $X$ is \rmi, $A_i'\coloneqq A_i\cup \{v_i\}$ fulfills the property $X$ for all $i\in [\ell - 1]$. Since $A_i$ can be transformed into $A_i'$ by a token addition step and $A_i'$ into $A_{i+1}$ by a token removal step, $A_s = A_1, A_1', A_2, \ldots, A_{\ell-1}, A_{\ell-1}', A_{\ell} = A_t$ is an $X$-TAR sequence of length $2\ell$. 

    Conversely, let  $A_1=B_1, \ldots, B_{\ell'} = A_t$ be an $X$-$\text{TAR}$ sequence of length $\ell'\leq 2\ell$. 
    We can assume that $B_i$ and $B_j$ are pairwise different for $i,j\in [\ell']$ if $i< j$. Otherwise,\longversion{ we can} delete the\longversion{ sets} $B_i, B_{i+1},\ldots, B_{j-1}$. 
    The resulting sequence would\longversion{ also} be an $X$-$\text{TAR}$ sequence of a length at most $2\ell$. Also, observe that $\ell'$ is odd.
    
    Assume there \shortversion{is}\longversion{exists} an $i\in [\ell']$ with $\vert B_i \vert < k$. Clearly, $i\notin\{1,\ell'\}$.
    Let\longversion{ now} $i\in [\ell'-1]\setminus \{1\}$\longversion{ be an index} such that $\vert B_i\vert $ is minimum\longversion{ with $\vert B_i \vert < k$}. We \shortversion{now}\longversion{want to} show that there is an $X$-$\text{TAR}$ sequence where we deleted $B_i$ or found a $B_i'$ \shortversion{to replace}\longversion{which can substitute} $B_i$\longversion{ in the sequence}, with $\vert B_i \vert < \vert B_i' \vert $. This reduces the number of sets of minimum cardinality in the \longversion{considered }sequence.
    
    As  $\vert B_i\vert $ is minimum, $B_{i-1}$ can be transformed into $B_{i}$ by a token removal step and $B_i$ can be transformed into $B_{i+1}$ by a token addition step. Hence, there \shortversion{is}\longversion{exists} a $ v\in B_{i-1}\setminus B_i$ and $u \in B_{i+1}\setminus B_i$. Thus, $B_{i+1} = (B_{i - 1} \setminus \{v\}) \cup \{u\}$. If $v=u$ (so $B_{i-1}=B_{i+1}$), we could delete $B_i$ and $B_{i+1}$\longversion{ from the sequence}. Otherwise, $B_{i-1}$ can be transformed into $B_{i+1}$ by a \longversion{token jumping}\shortversion{TJ} step. $B_i'\coloneqq B_{i-1} \cup \{u\}$ fulfills\longversion{ the property} $X$ as $X$ is \rmi. Hence, $B_1,\ldots,B_{i-1},B_i',B_{i+1},\ldots, B_{\ell'}$ is an $X$-$\text{TAR}$ sequence. We can \shortversion{repeat this}\longversion{do this iteratively}, until for each component of the sequence, the cardinality is at least~$k$. Let $C_1,C_2,\ldots, C_{\ell''}$ denote the \longversion{finally }obtained sequence. This is an $X$-$\text{TAR}$ sequence  with threshold $k+1$ of odd length $\ell''\leq \ell'$. By construction, \shortversion{TAR-}\longversion{token addition and token removal }steps always alternate in this sequence. Hence, $\vert C_i\vert = k + 1$  for all even $i \in [\ell'']$,  and $\vert C_i\vert = k$ for all odd $i \in [\ell'']$. As the components of the sequence are pairwise different, \shortversion{$B_1=C_1,C_3,\ldots,C_{\ell''}=B_{\ell'}$}\longversion{$C_1,C_3,\ldots,C_{\ell''}$} is an $X$-$\text{TJ}$ sequence\longversion{ with $C_1=B_1$ and $C_{\ell''}=B_{\ell'}$} of length $\left\lceil\frac{\ell''}{2}\right\rceil\leq \ell$.
\end{pf}

\begin{proposition}\label{prp:ReconfMonoAlliance}
    The properties 
    $X$ and $\textsc{G-}X$ are \rmi for \longversion{$X\in \{\,\textsc{Def-All}, \textsc{Off-All}, \textsc{Pow-All}\,\}$}\shortversion{$X\in \{\,\textsc{DA}, \textsc{OA}, \textsc{PA}\,\}$}. 
\end{proposition}

\begin{pf}
    Let $A,B\subseteq V$ be \defalls such that $A$ can be transformed into $B$ by a \shortversion{TJ}\longversion{token jumping} step with $u\in A\setminus B$ and $v\in B\setminus A$. Define $C\coloneqq A \cup \{ v \} = B \cup \{ u \}$. For $x\in A\subseteq C$, $d_C(x) + 1 \geq d_A(x) + 1 \geq d_{V\setminus A}(x) \geq d_{V\setminus C}(x)$. Further $d_C(v) \geq d_B(v) \geq d_{V\setminus B}(v) \geq d_{V\setminus C}(v)$. Hence, the property of being a \defall is \rmi. \shortversion{The proof for \textsc{OA} works analogously.} 
    
    \longversion{Next, we consider the property \textsc{Off}. Therefore, let $A,B\subseteq V$ be \offalls such that $A$ can be transformed into $B$ by a token jumping step. Here the token jumps from $u\in A\setminus B$ to $v\in B\setminus A$. Define $C\coloneqq A \cup \{ v \} = B \cup \{ u \}$. Let $x\in \partial C\subseteq (\partial A)\cup (\partial B)$. For $x\in \partial A$, $d_C(x) \geq d_A(x) \geq d_{V \setminus A}(x) \geq d_{V \setminus C}(x)$. If $x\in \partial B$, $d_C(x) \geq d_B(x) \geq d_{V \setminus B}(x) \geq d_{V \setminus C}(x)$. Hence, the property of being an \offall is \rmi. }

    A\shortversion{ \powall}\longversion{n alliance is powerful if it} is \longversion{both }defensive and offensive. By \autoref{pro:ReconfMonoCup}, the property \longversion{\textsc{Pow}}\shortversion{\textsc{PA}} is \longversion{also }\rmi. Since domination is a \longversion{monotone increasing}\shortversion{\rmi} property, $\textsc{G-}X$
    is \rmi for \longversion{$X\in \{\,\textsc{Def-All}, \textsc{Off-All}, \textsc{Pow-All}\,\}$}\shortversion{$X\in \{\,\textsc{DA}, \textsc{OA}, \textsc{PA}\,\}$}.
\end{pf}

\noindent
\longversion{The property} $\textsc{Idp-Off}$ is not \rmi. \shortversion{A token could jump to a neighbor, disabling}\longversion{It could be the case that a token jumps to a neighboring vertex. Thus, we cannot use} \autoref{thm:ReconfMonoTAR}. 

\begin{theorem}\shortversion{$(*)$}\label{thm:i_off_all_tar}
    Let $G=(V,E)$ be a graph and $A_s,A_t\subseteq V$ be independent \offalls \shortversion{s.t.}\longversion{with} $\vert A_s\vert = \vert A_t\vert = k$. There is an independent \offalltjumpseq \iffl there is an independent \offalltarseq from $A_s$ to $A_t$ with threshold $k+1$. 
\end{theorem}

\begin{toappendix}
\begin{pf}\shortversion{[of \autoref{thm:i_off_all_tar}]}
Let $A, B \subseteq V$ be independent \offalls with $k$ vertices such that $A$ can be transformed into $B$ by token jumping; say, $u\in A \setminus B$ and $v\in B\setminus A$. If $\{u,v\}\notin E$, then we can first insert~$v$ into~$A$, yielding a set $A'$, and then delete~$u$. As $A$ and $B$ are \offalls, $A'$ is an \offall by \autoref{prp:ReconfMonoAlliance}. As $B$ is an independent set, $v$ is not a neighbor of any vertex from~$A$ but possibly~$u$, but this is excluded in this case, so that $A'$ is also an independent set. For $\{u,v\}\in E$, \autoref{lem:i_off_ts_iso_edge} implies that $\{u,v\}$ is an isolated edge. Therefore, we can make changes on this edge independently of the other parts of the graph. Hence, $A,A\setminus \{u\},B$ is an independent \offalltarseq. By using this procedure repeatedly, this argument implies the only-if-direction.

For the if-direction, let $A_s=A'_1,\ldots,A'_{\ell}=A_t$ be an independent \offalltarseq with $\vert A'_i\vert \leq k+1$. If there exists an $i\in [\ell-2]$ and a $v\in V$ such that $v\in (A'_i\cap A'_{i+2})\triangle A'_{i+1}$, then we can delete $A'_{i+1},A'_{i+2}$ from the sequence, since $A'_i=A'_{i+2}$. Hence, we can assume this is not the case in the sequence. 

Assume there exists an $i\in [\ell-1]$ ($i\neq 1$) such that $u\in A'_{i-1}\setminus A'_i$ and $v\in A'_{i+1}\setminus A'_i$ with $u \neq v$ and $\{v,u\}\in E$. Then, $A'_{i-1}$ can be transformed into $A'_{i+1}$ by a token sliding step. By \autoref{lem:i_off_ts_iso_edge}, $\{u,v\}$ forms an isolated edge in~$G$. Therefore, these two transformations do not affect the other parts of the alliances. Furthermore, we can assume that $u\notin A'_{i-2}$, as otherwise this would contradict the independence of $A'_{i-2}$. By the argumentation above, we can assume that $v\in A'_{i-2}$. Then $A'_{i-2},A'_{i-2}\setminus \{v\},(A'_{i-2}\cup \{u\})\setminus \{v\},(A'_{i-1}\cup \{u\})\setminus \{v\}=A_{i+1}$ forms an independent \offalltarseq. Thus, such a step will be done at the beginning of our sequence. 
 
Choose $i\in [\ell]$ such that $\vert A'_i\vert$ is minimal. Assume $\vert A'_i\vert < k-1$. Since $\vert A'_i\vert$ is minimal, $\vert A'_{i-1}\vert=\vert A'_{i+1}\vert=\vert A'_i\vert +1$. Let $u\in A'_{i-1}\setminus A'_i$ and $v\in A'_{i+1}\setminus A'_i$. Because of the arguments above, $u \neq v$ and $\{v,u\}\notin E$. Define $\widetilde{A}_i:=A'_{i+1}\cup \{u\}=A'_{i-1}\cup \{v\}=A'_{i}\cup \{u,v\}$. We can use the same idea as in \autoref{thm:ReconfMonoTAR} to prove that $A_s=A'_1,\ldots, A'_{i-1}, \widetilde{A}_i,A'_{i+1},\ldots, A'_{\ell}=A_t$ is an \offalltarseq. This sequence is also independent, as $\{v,u\}\notin E$ and $\widetilde{A}_i \setminus \{u\}=A'_{i+1}$ and $\widetilde{A}_i \setminus \{v\}=A'_{i-1}$ are independent.

Therefore, we can assume we have an  independent \offalltarseq $A_s=A'_1,\ldots,A'_{\ell}=A_t$, such that  $\vert A'_i \vert = k $ for odd $i\in [\ell]$ and $\vert A'_i \vert \in \{k-1,k+1\} $ for even $i\in [\ell]$. So, $A_s=A'_1,A'_3,\ldots,A'_{\ell-2},A'_{\ell}=A_t$ is an independent \offalltjumpseq.
\end{pf}    
\end{toappendix}

\begin{lemma}\shortversion{$(*)$}
\label{lem:g_i_o-structure}
Let $G=(V,E)$ be graph. For two global independent \offalls $A_s,A_t \subseteq V$ of~$G$, there exists no global independent \offalltarseq. 
\end{lemma}

\begin{toappendix}
\begin{pf}\shortversion{[of \autoref{lem:g_i_o-structure}]}
    Let $A_S$ be global independent \offall. For any $v\in A_s$, $A_s\setminus \{v\}$ is not global anymore, as $A_s$ is independent. Since $A_s$ is a dominating set, for each $u \in V \setminus A_s$, there exists a $v\in A_s\cap N(u)$. Therefore,  any $A_s\cup \{u\}$ is not independent for any $u\in V\setminus A_s$. 
\end{pf}    
\end{toappendix}

For a reconfiguration problems $\Pi$-\textsc{Reconf}, the reconfiguration graph is often considered. In this graph, the vertices represent a set with the given property and a feasible size. The edges imply that the sets can be transformed into each other by one corresponding transformation step. The previous lemma implies that any reconfiguration graph for \textsc{G-Idp}-\offallianceReconf-TAR has no edges.
This has the following trivial algorithmic implication.

\begin{corollary}\label{cor:g_i_offensive_tar} \shortversion{$(*)$}
    \textsc{G-Idp}-\offallianceReconf-TAR can be solved in \textsf{LogSpace}.
\end{corollary}

\begin{toappendix}
\begin{pf}\shortversion{[of \autoref{cor:g_i_offensive_tar}]}
    The algorithm is just comparing the start with target reconfiguration. If they are the same, then it is a \yes-instance. Otherwise, this is a \no-instance.
\end{pf}    
\end{toappendix}

These results, together with \longversion{the results from }\shortversion{\autoref{thm:all_All_TJ}}\longversion{\autoref{subsec:tj_Pspace}}, imply a number of further \pspace-completeness results for TAR-reconfiguration problems, as summarized \shortversion{next}\longversion{in the following}.

\begin{corollary}\label{cor:all-reconf-TAR} For \longversion{$X\in\{$\textsc{\,Def},\textsc{ Off},\textsc{ G-Def}, \textsc{ G-Off}, \textsc{ Pow}, \textsc{ G-Pow},\textsc{ Idp-Off\,}$\}$,
    $X$-\AllReconf-TAR}\shortversion{$X\in\{\,\textsc{DA}, \textsc{OA}, \textsc{G-DA}, \textsc{G-OA}, \textsc{PA}, \textsc{G-PA}, \textsc{Idp-OA}\,\}$, $X$-\textsc{Reconf}-TAR} is \pspace-complete, even on bipartite graphs. \textsc{G-}\offallianceReconf-TAR is also \pspace-complete on chordal graphs.
\end{corollary}

\section{\FPT-algorithms: Natural Parameters and Limitations}
\label{sec:alliance-fpt}

In this section, we will show that there are \FPT-algorithms for \defallianceReconf-TJ/TS/TAR and \offallianceReconf-TS if the parameter is  the number of steps (denoted by~$\ell$) plus the cardinality of the alliances (denoted by~$k$).  The reader might wonder why we look at this combined parameter $k+\ell$. Notice that \pspace-hardness reductions are also \FPT-reductions with respect to $\ell$-\DSReconf-TJ and the corresponding alliance reconfiguration version parameterized by~$\ell$. By\longversion{ Mouawad et al.}~\cite{MouNRSS2017}, \longversion{it is known that }\DSReconf-TJ is \W{2}-hard if parameterized by~$\ell$.

\begin{corollary}\label{cor:W2_l_all_Reconf}\shortversion{$(\star)$} For \longversion{$ X\in \{\textsc{\,Def},\textsc{ Off}, \textsc{ Pow}, \textsc{ G-Def},\textsc{ G-Off}, \textsc{ G-Pow\,}\}$}\shortversion{$X\in \{\textsc{(G-)DA},\textsc{(G-)OA},\textsc{(G-)PA}\}$}\shortversion{,}\longversion{ and for} $Y\in\{\text{TS, TJ, TAR}\}$,
    $X$-\longversion{\AllReconf}\shortversion{\textsc{Reconf}}-$Y$ is \W{2}-hard  if parameterized by $\ell$; this also holds for \longversion{\textsc{Idp}-\offallianceReconf-TJ and \textsc{Idp}-\offallianceReconf-TAR}\shortversion{\textsc{Idp}-\offallianceReconf-TJ/TAR}. \shortversion{The corresponding}\longversion{All these} parameterized problems are in $XP$\longversion{ with this parameter}.
\end{corollary}
\begin{toappendix}
\begin{pf}(of \autoref{cor:W2_l_all_Reconf})
The \XP-algorithm is very simple and has been also observed in other contexts, see~\cite{MouNRSS2017}. As we have at most $\vert V\vert$ vertices, we can move a token from and to at most $\vert V\vert$ vertices. This leads to at most $\left( n^2 \right)^{\ell}= n^{2\ell}$ many reconfiguration sequences for \shortversion{TS}\longversion{token sliding} and \shortversion{TJ}\longversion{token jumping} that we need to consider. For TAR, we have only $(2 \cdot n)^{\ell}$ many sequences.
\end{pf}
\end{toappendix}
In each case, we will provide a\shortversion{n}\longversion{ function} $f: \mathbb{N}\to \mathbb{N}$ such that, in each reconfiguration step, there are at most $f(k)$ many choices to consider. \longversion{This implies that}\shortversion{Hence,} there are at most $f(k)^{\ell}$ many reconfiguration sequences to be taken into account. Since we can check in polynomial time\longversion{ (with respect to the input size)} if such a sequence is feasible, we get \FPT-algorithms.

To gain just the information that there exists an \FPT-algorithm for each case with this idea, it would be enough to consider \longversion{token jumping}\shortversion{TJ}, since each \longversion{token sliding}\shortversion{TS} step is also a \longversion{token jumping}\shortversion{TJ} step. Nevertheless, we will also present the function $f$ for the \longversion{token sliding}\shortversion{TS} cases, as $f$ will be significantly smaller\longversion{ (which is good for the running time)}. Furthermore, this could be helpful to understand the ideas for the \longversion{token jumping}\shortversion{TJ} cases. Therefore, we will start with the \longversion{token sliding}\shortversion{TS} algorithms.  

\begin{theorem}\label{thm:def-reconf-TS-k+ell}
    $(k+\ell)$-\defallianceReconf-TS, $(k+\ell)$-\offallianceReconf-TS $\in \FPT$.
\end{theorem}
\begin{pf}
Let $G=(V,E)$ be a graph. At first we consider \shortversion{TS}\longversion{token sliding} for \defalls. In each step, we can move one of the tokens to one \longversion{of its }neighbor\longversion{s} which is not in the alliance. Since $k\geq d_{A}(v)+ 1 \geq d_{A\setminus S}(v)$ holds for all \defalls $A\subseteq V$ with $\vert A \vert =k$ and $v\in A$, in each step, we only can put one token of $A$ ($k$ possibilities) on one of at most $k$ neighbors. \shortversion{Thus}\longversion{Therefore}, in each of the at most $\ell$ steps, we have $f(k) = k^2$ possibilities. Hence, there are $\sum_{i=1}^\ell\left( k^2 \right)^i\leq \left( k^2 \right)^{\ell+1}$ many possible reconfiguration sequences. 

We consider $(k+\ell)$-\offallianceReconf-TS next. Let $A,B\subseteq V$ be \offalls where $A$ can be transformed into $B$ by one sliding step. Hence, there are  unique $v\in A\setminus B$ and $u\in B\setminus A$\longversion{. As $A$ can be transformed into $B$ by one token sliding step,}\shortversion{ with} $v\in N(u)$. 
Thus, $ d_{ V \setminus B}(v) = d_{ V \setminus A}(v) + 1$. As $B$ is an \offall, $ k \geq  d_B(v) \geq d_{ V \setminus B}(v) = d_{ V \setminus A}(v) + 1$. 
Hence for each token which we can move, there are at most $k-1$ possible new positions. As we have $k$ tokens, there are at most $f(k) = k^2-k$ possibilities of continuation in the next step. Hence, there are at most $\left( k^2-k\right)^{\ell+1}$\longversion{ many possible} reconfiguration sequences.
\end{pf}

\begin{toappendix}
\underline{Improvements of the algorithm in the proof of \autoref{thm:def-reconf-TS-k+ell}}.
The algorithms can be improved by `working from both ends', at the expense of admitting exponential-space space: the algorithm would first work (at most) $\ell/2$ steps from $A_s$ (storing all reached sets in a linear ordering) and then work backwards  (at most) $\ell/2$ steps from $A_t$ (storing all reached sets in a linear ordering); the sorting allows us to check if both lists of sets contain a common element. Hence, we only need to consider at most $2\sum_{i=1}^{\ell/2}\left( k^2-k\right)^{i}\leq \left( k^2-k\right)^{\ell/2+1}$ many reconfiguration sequences and check in the end if there are any alliances that are reachable from both given alliances by sliding at most $\ell/2$ tokens. (Here, one has to be careful if $\ell$ is odd, but these minor details can be fixed.)
\end{toappendix}

\longversion{The arguments leading to these algorithms already imply the same results for the powerful and global versions, as we only have to check after each step \shortversion{some additional properties}\longversion{if the current set is a \powall or / and a dominating set}. 

\begin{corollary}\label{Cor_FPT_TS}
    $(k+\ell)$-\textsc{G-}\defallianceReconf-TS, $(k+\ell)$-\textsc{G-}\offallianceReconf-TS, $(k+\ell)$-\longversion{\textsc{Pow}-\AllReconf}\shortversion{\textsc{(G-)PA-Reconf}}-TS\longversion{, and $(k+\ell)$-\textsc{G-Pow}-\AllReconf-TS} are in  $\FPT$.
\end{corollary}}

Now, we will consider the token addition/removal and jumping versions. We mostly prove the results directly for token addition/removal. Because of  reconfiguration monotonicity,  some results transfer directly to token jumping.

For $G=(V,E)$, let $G^{\leq r}=G[V^{\leq r}]$ with $V^{\leq r}=\{v\in V\mid d_G(v)\leq r\}$. If~$G'$ is a subgraph of~$G$, then let $\text{dist}_{G'}(x,y)$ denote the shortest-path distance within~$G'$, and $\text{dist}_{G'}(x,M)=\min\{\text{dist}_{G'}(x,y)\mid y\in M\}$. 
The following lemma decreases the number of TAR sequences we need to consider for \defallianceReconf-TAR. We use\longversion{ the fact} that each \defall of size at most~$k$ in~$G$ needs to be a subset of $V^{\leq 2k}$. \longversion{Furthermore}\shortversion{Also}, we can add at most $\ell$ tokens. \longversion{Therefore}\shortversion{Thus}, we only need to consider vertices in $v \in V^{\leq 2k}$, within a distance of $\ell$ to $A_s \cup A_t$ on $G^{\leq k}$, as the other vertices are irrelevant.

\begin{lemma}\label{lem:DA-distance}
Let $G=(V,E)$ be a graph and $A_s,A_t \subseteq V$ be \defalls, $\vert A_s\vert=\vert A_t\vert=k$, with a \defall TAR sequence $A_s=A_1,\ldots,A_{\ell}=A_t$ of length~$\ell$. 
Then there exists a \defall TAR sequence $A_s=A'_1,\ldots,A'_{\ell'}=A_t$,   with $\ell' \leq \ell$, such that $$\bigcup_{i=1}^{\ell'}A_i'\subseteq\{v\in V^{\leq 2k}\mid \text{dist}_{G^{\leq 2k}}(v,A_s\cup A_t)\leq \ell\}\,.$$
\end{lemma}
\begin{pf}
Clearly, we can assume that the mapping $[\ell]\to 2^V$, $i\mapsto A_i$ is injective.    We simplify the notation by setting $A \coloneqq \bigcup_{i=1}^{\ell}A_i$. 
Assume there is a $v\in A$ with $\text{dist}_{G^{\leq 2k}}(v,A_s \cup A_t) > \ell$\longversion{; otherwise, the statement is trivially satisfied}. 
    Define $K_j\subseteq A$ for $j\in [\ell] \cup \{0\}$ inductively by $K_0=\{v\}$ and 
    $K_j=K_{j-1} \cup K_j'$ for $j\in[\ell]$ with $$K_j'\coloneqq\{x\in A\setminus K_{j-1}\mid \exists i\in [\ell]: d_{A_i \setminus K_{j-1}}(x) + 1< d_{V \setminus (A_i \setminus K_{j-1})}(x)\}.$$
    
    In other words, $K_j'$ includes vertices for which there exists an $i\in [\ell]$ such that  $v$ violates the \defall property for $A_i \setminus K_{j-1}$.
    
    Let $j\in [\ell]$ be fixed. For $x \in V \setminus N[K_j]$, $d_{A_i \setminus K_{j-1}}(x) = d_{A_i}(x)$ as well as 
    $d_{V \setminus (A_i \setminus K_{j-1})}(x) =d_{V \setminus A_i}(x)$. Hence, $K_j \subseteq N(K_{j-1})$. By an inductive argument, $\text{dist}_{G^{\leq 2k}}(v,K_j)\leq j\leq \ell< \text{dist}_{G^{\leq 2k}}(v,A_s\cup A_t) $. So, $K_j\cap (A_s\cup A_t)=\emptyset$ which implies $A_s \setminus K_j = A_s$ and $A_t \setminus K_j = A_t$.
        
    Assume $K_j \neq K_{j-1}$ for all $j\in [\ell]$. Since $K_j$ is strictly monotone increasing, $\vert K_{\ell} \vert > \ell$. As $K_{\ell}\cap (A_s\cup A_t)= \emptyset$ and $K_\ell\subseteq A$, this contradicts\longversion{ the fact} that $A_s=A_1,\ldots,A_{\ell}=A_t$ is a TAR sequence, as only one vertex can be added per step.
    
    Hence, we can assume there is a $j\in [\ell]$ such that $K_{j-1}= K_j$, so $K'_{j}=\emptyset$. Thus, for all $q \in  [\ell] \setminus [j-1]$, $K_j=K_q$ and $K'_{q}=\emptyset$. By the definition of $K_j'$, $A_1\setminus K_j,\ldots,A_{\ell}\setminus K_j$ are \defalls. Furthermore, for each $i\in [\ell-1]$, $A_i\setminus K_j=A_{i+1}\setminus K_j$ or $A_i\setminus K_j$ can be transformed into $A_{i+1}\setminus K_j$ by a \shortversion{TAR}\longversion{token addition or removal} step. Hence, we can shorten the  $A_1\setminus K_j,\ldots,A_{\ell}\setminus K_j$ into a \defall TAR reconfiguration sequence $A_1',\ldots,A'_{\ell'}$ by deleting all but one sets \longversion{in the sequence }that are the same. Thus, for each $i\in [\ell']$, $v\notin A_i'$. 
    We can use this argument repeatedly to prove the lemma.
\end{pf} 


\autoref{lem:DA-distance} restricts the search space to $\{v\in V^{\leq 2k}\mid \text{dist}_{G^{\leq 2k}}(v,A_s\cup A_t)\leq \ell\}$, which gives (together with \autoref{thm:ReconfMonoTAR} and \autoref{prp:ReconfMonoAlliance}) the next result. 

\begin{theorem}\label{thm:def-TJ+TAR-param}
    $(k+\ell)$-\defallianceReconf-TAR, $(k+\ell)$-\defallianceReconf-TJ $\in \FPT$.
\end{theorem}

Before considering \offallianceReconf-TJ versions, we introduce an auxiliary result. To simplify the notation, we define for a graph $G=(V,E)$ and a set $X\subseteq V$, $Z(X)\coloneqq \{v\in V \setminus N[X] \mid N(v)\subseteq L \cup \partial X\}$ and $Y(X) \coloneqq N[X] \cup Z(X) \cup L$, where $L\coloneqq \{x\in V \mid d_G(x)=1\}$ is the set of leaves.
If $X$ is an \offall, $Z(X)$ are vertices $z \in N[X]$ for  which $X\cup \{z\}$ is still an \offall. \shortversion{Namely, the elements in $\partial X$ already fulfill the degree property for the \offalls and leaves fulfill this property if the neighbor is in $X$.} 
Clearly, $Z(X)$ is an independent set of $G$ for any\longversion{ set} $X\subseteq V$. 
$Y(X)$ is the union of the closed neighborhood of $X$ together with the leaves and $Z(X)$. The next lemma gives a combinatorial restriction on our search space. 
\begin{lemma}\label{lem:off_all_reconf_tar_boundary}
Let $G=(V,E)$ be a graph and $A,B$ be \offalls of~$G$ for which there \longversion{exists}\shortversion{is} an \offall TM sequence between both for $\text{TM}\in \{$\,TAR,\,TJ\,$\}$. Then, $Y(A)=Y(B)$.
\end{lemma}


\begin{pf}
By \autoref{thm:ReconfMonoTAR} and \autoref{prp:ReconfMonoAlliance} each TJ step can be seen as two TAR steps. So, it is enough to show that this lemma holds if  there \shortversion{is}\longversion{exists} a $v\in A$ with  $A= B \cup \{v\}$\longversion{, as this shows it for one TAR step; then an induction proves the lemma}\shortversion{. Using this argument inductively proves the lemma}. 

Since $A$ is an \offall, for each $x\in N(v)$, $x\in \partial A$ with $d_{A}(x)>d_{V \setminus A}(x)$ or $x\in A$.  Thus, if $x\in N(v)\setminus \left(L \cup A\right) = N(v)\setminus \left(L \cup B\right)$,  $(A \setminus \{v\})\cap N(x) = B \cap N(x)$ is not empty. Thus, $x\in \partial B $ and $N(v) \subseteq N[B] \cup L \subseteq Y(B)$. Therefore, $\partial A\subseteq Y(B)$.\longversion{

} Now, we want to show $v\in Y(B)$. If $v$ has a neighbor in $B$, then $v\in Y(B)$. If $v$ has no neighbor in $B$, then by the same argument as above, each vertex in $N(v)\setminus \left(L\cup A\right) = N(v)\setminus L$ has a neigbor in~$B$. Thus, $v \in Z(B)\subseteq Y(B)$.

This leaves to show $Z(A) \subseteq Y(B)$. Assume there exists a $u\in Z(A)$ such that $u \in Z(A)\setminus Y(B)$. As $u\in Z(A)\subseteq V\setminus N[A] \subseteq V \setminus N[B]$, $u \notin N[B]= B\cup \partial B$. Thus, $u$ has a neighbor $w \in (\partial A) \setminus L$ with $w\notin \partial B$. Therefore, $w\in (N[v] \cap N[u]) \setminus N[B]$. This is a contradiction to $N(v) \subseteq N[B] \cup L$ (see above). Hence, $Y(A)\subseteq Y(B)$. 

As $N[B]\subseteq N[A]$, \longversion{we only need to show that }$Z(B) \subseteq Y(A)$\shortversion{ has to be proved}. Let $u\in Z(B)$. Then, \longversion{either }$u\in N[A]\subseteq Y(A)$ or $N(u) \subseteq L \cup \partial B \subseteq L\cup N[A]$ (\shortversion{i.e., $u\in\partial A\cup Z(A)$}\longversion{thus, $u\in\partial A$ or $u\in Z(A)$, respectively, and hence $u\in Y(A)$}). 
\end{pf}

\shortversion{As detailed in the appendix, \autoref{lem:off_all_reconf_tar_boundary} can be helpful to get \FPT-algorithms for some versions of \offallianceReconf-TJ. 
We discuss another application  next.}
\begin{toappendix}
\begin{remark} \autoref{lem:off_all_reconf_tar_boundary} can be helpful to provide an \FPT-algorithm for some versions of \offallianceReconf-TJ. If we have a parameter $p$ that is an upper bound on the number of tokens and we can bound the number of vertices in $\partial A_s$ by $f(p)$ (where $f:\mathbb{N}\to \mathbb{N}$ is a computable function), then we can bound the number of vertices in $Z(A_s)\cap N(\partial A_s)$ by $p\cdot f(p)$. The remaining vertices in $Z(A_s)$ belong to stars. Since these stars do not intersect $A_s$ and it does not make a difference at which vertex in these stars we put a token, we need to consider at most $p$ of them and can delete the remaining ones.
We can consider each vertex in $Z(A_s)$ has at most $2p + 1$ leaf neighbors, as the others are redundant and can be deleted. These observations bound the number of the vertices the tokens can be at in $p$ (say $g(p)$). Therefore, there are at most $g(p)^p$ many \offalls we need to consider. 
For instance, if we parameterize \offallianceReconf-TJ both by $k$ and by the maximum degree of the graph, this reasoning yields an \FPT-result. For the parameter $k$ alone we cannot use this technique yet as $\partial A_s$ can be very large. 
\end{remark}   
\end{toappendix}
\longversion{\noindent We discuss another application of this lemma next.}
\begin{theorem}
$k$-(\textsc{G}-)\shortversion{PA-Reconf}$\longversion{\textsc{Pow-}\AllReconf}\text{-$Y$}\in \FPT$ for $Y\in\{\,\text{TJ, TS, TAR}\,\}$. 
\end{theorem}

\begin{pf}
Let $G=(V,E)$ be a graph and let $A_s=A_1,\ldots,A_{\ell} = A_t \subseteq V$ be a \powalls \shortversion{TAR}\longversion{token addition removal} sequence with $\vert A_i\vert \leq k$, for $i\in [\ell]$.  By \autoref{lem:off_all_reconf_tar_boundary}, for all $i \in [\ell]$, $Y(A_s)=Y(A_i)$. 
As $A_s$ is a \defall, $\vert \partial A_s \vert \leq k^2$. Furthermore, $\vert N(\partial A_s) \setminus A_s \vert \leq k^3$, as $A_s$ is an \offall. \longversion{

}
Consider the case that there is an $i\in [\ell-1]$ such that $A_i$ can be transformed into $A_{i+1}$ by adding a token to $x\in V\setminus N[\partial A_i]$. As $A_{i+1}$ is a \defall, $d_G(x)\leq 1$. If $x$ has a neighbor $u$, then $d_G(u)=1$, i.e., we have an isolated edge $\{x,u\}$. Otherwise, $A_{i+1}$ is not an \offall. Let $M$ be the set of such vertices. 
The vertices in $M$ can be also useful for \longversion{token addition}\shortversion{TA} steps, but clearly, we only need to consider at most~$3k$ of them for reconfiguration, as there is no  difference for the sequence differentiating on which vertex in $M\setminus (A_s\cup A_t)$ we put a token: collect these in $M_k\subseteq M$. Define $Y\coloneqq Y(A_s)\setminus (M\setminus M_k)$. Hence, there are at most $|Y|\leq k^2+k^3+3k$ vertices which are useful for any reconfiguration step. Thus, the number of  \powalls which are useful for the reconfiguration sequence is at most $r_k=\sum_{j=0}^k\binom{k^2+k^3+3k}{j}$. This number is also an upper bound on the number of steps as we can avoid visiting sets twice in the sequence. Therefore, the algorithm runs in 
$\mathcal{O}^*\left(\left(k^2+k^3+3k\right)^{r_k}\right)$ time.\longversion{

}
\shortversion{As these are \rmi properties}\longversion{By using reconfiguration monotonicity}, we also get \FPT-results for \longversion{toking jumping and sliding}\shortversion{TJ and TS}. Additionally  checking if the alliances are dominating sets yields \FPT-algorithms for the \longversion{global }\shortversion{\textsc{G}-}variants.
\end{pf}

Quite similarly, one can also attack other alliance reconfiguration problems, even only with the single  parameter~$k$, as we can bound the number of steps of a reconfiguration sequence by a computable function in~$k$.

\begin{theorem}\shortversion{$(*)$}\label{thm:g_o_k}
$k$-$\textsc{G-}\offallianceReconf\text{-$Y$}\in \FPT$ for $Y\in\{\,\text{TJ, TS, TAR}\,\}$. 
\end{theorem}
\begin{toappendix}
\begin{pf}\shortversion{[of \autoref{thm:g_o_k}]}
    Let $G=(V,E)$ be a graph and $A_s,A_t\subseteq V$ be global \offalls with $k \coloneqq \vert A_s\vert= \vert A_t\vert$. Since $A_s$ is global, $V= A_s\cup \partial A_s$. Let $v\in V$ with $d_G(v)>2k$, then $v$ has to be in any global \offall $A$. Otherwise, $v \in \partial A$ and $d_{V}(v) \leq k < d_{V \setminus A}(v)$, would contradict that $A$ is an \offall. Therefore, $D \coloneqq \{v\in V\mid d_G(v) > 2k\} \subseteq A$. Define $B = \{v\in V \setminus D \mid d_{D}(v) > d_{V \setminus D}(v)\}$. We need not check $d_A(v) \geq d_{V \setminus A}(v)$ for $v\in (\partial A)\cap B $ to verify that $A \subseteq V$ with $\vert A\vert =k$ is an \offall, since $D\subseteq A$ has to hold for such an \offall. For $v\in \left( \partial A_s \right) \setminus B$, there has to be a $u\in A_s \setminus D$. Since for all $u\in A_s\setminus D$, $d_G(u)\leq 2k$, there are at most $2k^2$ many vertices in $\left(\partial A_s \right)\setminus B$. For $W\subseteq V' \coloneqq V\setminus \left( D \cup B\right)$, define $P_W\coloneqq \{v\in B \mid N_{V'}(v)=W\}$. Clearly, $\mathcal{P}=\{P_W \mid W\subseteq V'\}$ is a partition of $B$ in to $2^{k+2k^2}$ classes. The following claim will help us to bound the number of global \offalls that we need to consider in our sequences.

    \begin{claim}
    Let the following be given: $W\subseteq V'$, $v\in P_W$, a global \offall $A\subseteq V \setminus \{v\}$ of~$G$ with $\vert A\vert =k$ and $u \in A\cap P_W \neq \emptyset$. Then $A'\coloneqq \left( A\cup \{v\}\right) \setminus \{u\}$ is also a global \offall  of~$G$. 
    \end{claim}
    \begin{pfclaim}
        Since $A$ is a global alliance for cardinality $k$, $D \subseteq A$. Let $w\in V\setminus A'$. If $w\in B$, then $w \in \partial A'$, by definition of $B$. For $w \in V' \setminus W$, $w$ has to have a neighbor in $A\setminus \{u\} = A' \setminus \{v\}$, otherwise $A$ would not be a global \offall. By definition of $P_W$, $v\in N[w] \cap A'$, if $w\in W$. Hence, $A'$ is a dominating set. 
        
        This leaves to show that $A'$ is an \offall. Let $w \in \partial A'$. For $w\in B$, $D\subseteq A'$ and the definition of $B $ imply $d_{A'}(w) > d_{V \setminus A'}(w)$. For each $w\in \left( \left( \partial A'\right) \cap V'\right) \setminus W$, $N_A(w)=N_{A'}(w)$ and $N_{V \setminus A}(w)=N_{V \setminus A'}(w)$. Therefore, we can assume $w\in \left( \partial A'\right) \cap  W$. Since $N_A(w)=\left(N_{A'}(w)\setminus \{u\} \right) \cup \{ v \}$ and $N_{V \setminus A}(w)=\left(N_{V \setminus A'}(w)\setminus \{v\} \right) \cup \{ u \}$, $d_{A'}(w) = d_{A}(w) > d_{V \setminus A}(w) = d_{V \setminus A'}(w)$. Therefore, $A'$ is a global \offall with $\vert A'\vert = k$.  
    \end{pfclaim}
    By the claim, we now that we need to consider at most $2k$ vertices per partition class ($2k$ as $A_s$ and $A_t$ could have many vertices in one class). We can use this observation as a reduction rule. This implies $\vert \bigcup_{W \subseteq V'}P_W\vert\leq 2k\cdot 2^{k+2k^2}$ and $\vert V \vert\leq 2k\cdot 2^{k+2k^2} + k+2k^2$. So, there are at most $\binom{2k\cdot 2^{k+2k^2} + k+2k^2}{k}$ many global \offalls that we need to consider and we can assume that a reconfiguration sequence would have at most this length to avoid repetitions.
\end{pf}     
\end{toappendix}

\longversion{The problem $k$-$\textsc{G-}\defallianceReconf$-TJ can also be solved in \FPT-time, and so can the TS- and TAR-variants.}

\begin{theorem}\label{thm:g_def-TJTSTAR-FPT}\shortversion{$(*)$}
$k$-$\textsc{G-}\defallianceReconf\text{-$Y$}\in \FPT$ for $Y\in\{\,\text{TJ, TS, TAR}\,\}$. 
\end{theorem}
\begin{toappendix}
\begin{pf}\shortversion{[of \autoref{thm:g_def-TJTSTAR-FPT}]}
Let $G=(V,E)$ be a graph and $A_s, A_t \subseteq V$ be global \defalls with $k \coloneqq \vert A_s\vert = \vert A_t \vert  $. Since $A_s$ is global, $V= A_s\cup \partial A_s$. Since for each $v\in A_s$, $k = \vert A_s\vert \geq d_{A_s}(v) + 1 \geq d_{V\setminus A_s}(v)$, we can assume $\vert \partial A_s \vert \leq k^2$. Thus, $\vert V\vert \leq \vert A_s\vert + \vert \partial A_s\vert \leq k + k^2.$ Therefore, $G$ can have at most $\ell_k=\binom{k^2+k}{k}$ many global \defalls of cardinality $k$. Hence, there are at most this many steps in a global alliance reconfiguration token jumping (resp. sliding) sequence, such that there is no global \defall which appears twice in this sequence. So there are at most $\sum_{i=1}^{\ell_k}l_k^i$ many sequences that we need to consider.

For $k$-$\textsc{G-}\defallianceReconf$-TAR, the argumentation is analogous. We only need to keep in mind that there can be also global \defalls with less than $k$ vertices. Therefore, there are at most $l_k'\coloneqq\sum_{j=0}^k\binom{k+k^2}{j}\leq 2^{k+k^2}$ many global \defalls and 
$\sum_{i=1}^{l_k'}\left(l_k'\right)^i$
many global defensive TAR  reconfiguration sequences we need to consider.
\end{pf}    
\end{toappendix}


These are interesting results as \cite{BodGroSwe2021a}\longversion{\footnote{A short version appeared in \cite{BodGroSwe2021}.}} shows that  $k$-\DSReconf-TJ is \XL-complete, $k$-T-\DSReconf-TJ 
is \XNL-complete if the maximal number~$\ell$ of steps is given in binary (see \cite[Cor. 28]{BodGroSwe2021a}) and \XNLP-complete if $\ell$ is given in unary (see \cite[Theorem 36, Corollary 37]{BodGroSwe2021a}) if parameterized by the cardinality~$k$ of the dominating sets. For the definition of these classes, we refer to \cite{BodGroSwe2021a}.
By definition, $\XL\subseteq \XNL$. Chen and Flum~\cite{CheFlu2003a} proved $\XNLP\subseteq \XNL$. By Bodlaender \emph{et al.}~\cite{BodGNS2021},  \XNLP-hardness implies \W{t}-hardness for each $t\in \mathbb{N}\setminus \{0\}$. Our results on the reconfiguration of  global \longversion{defensive or offensive }alliances differ from the results for \DSReconf.

\section{Neighborhood Diversity}
\label{sec:alliance-nb-diversity}

We now consider the structural parameter  neighborhood diversity $\nd$ for our reconfiguration problems.
 It is known that if the vertex cover number~\textsf{vc} \longversion{or the parameter `distance to clique' }is upper-bounded by~$d$, then this also holds for the neighborhood diversity. Thus, an \FPT-algorithm with respect to  $\nd$ would also imply an \FPT-algorithm \longversion{with respect to}\shortversion{for}~\textsf{vc}. 

\begin{observation}\label{obser_nd}
    Let $G=(V,E)$ be a graph, $A$ be a \defall and $v,u$ be vertices of the same type. If $v\in A$ and $u\notin A$, then $\left( A\setminus\{v\} \right)\cup \{u\}$ is also a \defall.  Similar statements hold for \shortversion{(G-)}\offalls, \shortversion{(G-)}\powalls, global \defalls, \longversion{global \offalls, global \powalls,} and independent \offalls. 
\end{observation}

\begin{theorem}
    $\left(\nd+p\right)$-\textsc{Z}-\longversion{\AllReconf}\shortversion{\textsc{Reconf}}-TM $\in\FPT$ for TM${}\in \{ \text{TJ, TS, TAR}\}$, $p\in \{k,\ell\}$ and $\textsc{Z}\in\{$ \longversion{\textsc{Def, G-Def, Off, G-Off, Idp-Off,\linebreak[3] Pow,\linebreak[3] G-Pow}}\shortversion{\textsc{(G-)DA, (G)-OA, (G)-PA, Idp-OA}}$\}$.
\end{theorem}
\begin{pf}
Let $G=(V,E)$ be a graph, $k\in \mathbb{N}$ and $A_s,A_t \subseteq V$ be \defalls with $\vert A_s \vert = \vert A_t \vert =k$. Further, $d\coloneqq \nd(G)$ and let $C_1,\ldots,C_d$ be the neighborhood diversity equivalence classes. By \autoref{obser_nd}, we can assume that it is unimportant to which vertex in a class a token moves (unless \longversion{the vertex is}\shortversion{being} in $A_s$ or~$A_t$).

\longversion{We first}\shortversion{First, } consider $p=k$. We only need $2k$ vertices per class to remember. Hence, we\longversion{ only need $d\cdot 2k$ vertices 
and} have at most $\binom{d\cdot 2k}{k}$ many possible \defalls. So, we \shortversion{can}\longversion{only have to} go through all possibilities, check these and find a shortest path \longversion{through}\shortversion{in} this part of the reconfiguration graph.

Now we consider $p = \ell$. We move the token to a vertex in $A_t$, if possible. In the other cases, it is arbitrary to which vertex we move the token. Hence, there are at most $d^2$ many possible moves in one token transformation. Thus, there are at most $d^{2 \ell}$ many alliance reconfiguration sequences that we need to consider. 
This argument works analogously for the other versions of alliances. 
\end{pf}

Beside the sketched combinatorial algorithm of the last proof, we could also use Integer Linear Programming (ILP\longversion{ for short}) to solve these parameterized problems in \FPT-time. Using ILP for solving reconfiguration problems appear to be a new approach.\longversion{\footnote{In a completely different way, Ringel studied ILPs in the context of reconfiguration in \cite{Rin2024}.}}\shortversion{ Details are in the appendix.} It is still \underline{open} if we can get \FPT-results if we parameterize \longversion{the }alliance reconfiguration problems by $\nd$ only. \longversion{Notice that w}\shortversion{W}e cannot employ the meta-theorem from~\cite{GimIKO2022}: alliance problems are not expressible in MSO.

\begin{toappendix}
\noindent\underline{Bringing ILPs into the game.}
First we will fix our notation. Let $G=(V,E)$ be a graph with the neighborhood diversity classes $C_1,\ldots,C_{nd(G)}$. For $i \in [\nd(G)]$, $N_i\coloneqq \{j\in [\nd(G)] \mid C_j\cap N(C_i)\neq \emptyset \}$, $d_i$ denotes the degree of a vertex in $C_i$ and $c_i\in \{0,1\}$ is 1 \iffl $C_i$ is a clique. 
 For the ILP we need the variables, $x_{i,p}\in \{0,\ldots, \vert C_i\vert \}, \, y_{i,j,p}\in \{0,1\}$ for $i,j\in [\nd(G)]$ and $p\in [\ell - 1]$. $x_{i,p}$ tells how many tokens are in $A_{p} \cap C_i$. $y_{i,j,p}$ is 1 \iffl a token jumps from $C_i$ to $C_j$ in the transformation between $A_p$ and $A_{p+1}$. 

Now, we give the (in)equalities that a feasible solution of any alliance reconfiguration problem should satisfy:
\small
\begin{align}
        \sum_{i,j=1}^{\nd(G)} y_{i,j,p}&\leq 1 & \forall p\in [\ell - 1] \label{con:one_step}\\
        x_{i,1}& =\vert C_i\cap A_s\vert  & \forall i\in [\nd(G)]\label{con:start}\\
        x_{i,p+1} &= x_{i,p}-\left(\sum_{j=1}^{\nd(G)} y_{i,j,p} - y_{j,i,p}\right)  &\forall i\in [\nd(G)], p\in [\ell-2]\label{con:between}\\
        \vert C_i\cap A_t \vert &= x_{i,\ell-1}-\left(\sum_{j=1}^{\nd(G)} y_{i,j,\ell-1} - y_{j,i,\ell-1}\right)  &\forall i\in [\nd(G)]\label{con:end}\\ 
        \vert (C_i\cap A_s )\setminus A_t\vert &\leq\left(\sum_{p=1}^{\ell - 1}\sum_{j=1}^{\nd(G)} y_{i,j,p}\right)  &\forall i\in [\nd(G)]\label{con:sortClass}
\end{align}
\normalsize
The first inequality ensures that we do not make two jumps at the same time. (\ref{con:start}) and (\ref{con:end}) verify that $A_s$ is the start configuration and $A_t$ the end configuration. By (\ref{con:between}), the steps implied by $y_{i,j,p}$ fit with the alliances.  (\ref{con:sortClass}) ensures that the tokens on the vertices in $(C_i\cap A_s )\setminus A_t$ are no longer in this set at the end.

If we consider a version that needs the \defall property, we add the variable $w_{i,p}\in \{0,1\}$ for $i\in [\nd(G)]$ and $p\in [\ell - 1]$. $w_{i,p}=1$ holds \iffl $x_{i,p}\neq 0$. To verify these properties, we need the following inequalities:
\small
\begin{align}
         w_{i,p} d_i&\leq 2\cdot\left(-c_i \cdot w_{i,p}  + \sum_{j\in N_i}x_{j,p}\right) &\forall i\in [\nd(G)], p\in [\ell-1]\label{con:def_all}\\
        x_{i,p} &\leq \vert V\vert \cdot w_{i,p} &\forall i\in [\nd(G)], p\in [\ell-1]\label{con:w1if}\\
        w_{i,p} &\leq  x_{i,p} &\forall i\in [\nd(G)], p\in [\ell-1]\label{con:w0if}
\end{align}
\normalsize
(\ref{con:def_all}) ensures the \defall property. By (\ref{con:w1if}) and (\ref{con:w0if}), it is known that $w_{i,p}=1$ \iffl $x_{i_p} \neq 0 $. Hence, the left-hand side of (\ref{con:def_all}) is $0$ if $x_{i,p}$ is. In this case, the right-hand side is at least 0. Therefore, if $A_p\cap C_i$ is empty, then the inequality holds. Since $d_A(v)\geq d_{V\setminus A}(v) = d(v)-d_{A}(v)$ holds for each \defall, this inequality ensures that $A_p$ is a \defall. We need the~$c_i$ as we do not want the count the vertex itself for the degree.

For the \offall property, we add the variables $w'_{i,p},w''_{i,p}\in \{0,1\}$ for $i\in [\nd(G)]$ and $p\in [\ell - 1]$. $w'_{i,p}$ will be 1 \iffl a neighbor vertex of a vertex in $C_i$ is in $A_p$. $C_i\subseteq A_p$ \iffl $w''_{i,p}=0$. We add the following inequalities: 
\small
\begin{align}
        ( w'_{i,p} + w''_{i,p} - 1) (d_i + 2c_i + 1) & \leq 2\cdot\left(\sum_{j\in N_i}x_{j,p}\right) &\forall i\in [\nd(G)], p\in [\ell-1]\label{con:off_all}\\
        \sum_{j \in N_i} x_{j,p} &\leq \vert V\vert \cdot w'_{i,p} &\forall i\in [\nd(G)], p\in [\ell-1]\label{con:w'1if}\\
        w'_{i,p} &\leq  \sum_{j \in N_i} x_{j,p} &\forall i\in [\nd(G)], p\in [\ell-1]\label{con:w'0if}\\
        \vert C_i\vert - x_{i,p} &\leq \vert V\vert \cdot w''_{i,p} &\forall i\in [\nd(G)], p\in [\ell-1]\label{con:w''1if}\\
        w''_{i,p} &\leq \vert C_i\vert - x_{i,p} &\forall i\in [\nd(G)], p\in [\ell-1]\label{con:w''0if}
\end{align}
\normalsize
The inequalities (\ref{con:w'1if}) to (\ref{con:w''0if}) ensure the conditions that we require for $w'_{i,p},w''_{i,p}$. Inequality~\eqref{con:off_all} implies the \offall property. The left-hand side of this inequality is only larger than~$0$ if $w'_{i,p}=w''_{i,p} = 1$ while this holds in each case for the right-hand side. Hence, we only need to consider this inequality if $C_i\nsubseteq A_p$ and a neighbor of the vertex in $C_i \setminus A_p$ is in $A_p$. In this case, these inequalities hold \iffl the set $A_p$ is an \offall for each $p\in[\ell - 1]$.

For global versions, we need the following additional inequalities:
\small 
\begin{align}
         1 &\leq \sum_{j\in N_i}x_{j,p} &\forall i\in [\nd(G)], p\in [\ell-1].\label{con:ILP_global}        
\end{align}
\normalsize
The independence property is ensured by 
\small
\begin{align}
         \vert V \vert  \cdot \sum_{j \in N_i}(x_{j,p} -c_i) &\leq x_{i,p} &\forall i\in [\nd(G)], p\in [\ell-1]\,.\label{con:ILP_indpendent}
\end{align}
\normalsize
In such an ILP, there are at most $ (\ell-1) \nd(G)( \nd(G) + 4)$ many variables and $(\ell-1)  + \ell \nd(G)+ \nd(G)+ 10 \nd(G)(\ell-1) $ many (in)equalities. Then~\cite[Theorem 6.5]{CygFKLMPPS2015} gives an \FPT-algorithm.    
\end{toappendix}

\section{Conclusions}

\begin{table}[t]\centering
\begin{tabular}{l||l|l|l|l|l|l|l|l}
       & \sc Def & \sc Off &\sc pow &\sc Idp-Off &\sc G-Def &\sc G-Off &\sc G-Pow &\sc G-Idp-Off \\\hline\hline
       TS & \ref{thm:defall_TS_PSpace} 
       & \shortversion{\ref{thm:all_All_TJ}}\longversion{\ref{thm:offall_TS_PSpace}}
       & \ref{cor:global-TS} & (\ref{prp:ind_off_ts_logspace})
       & \ref{cor:global-TS} & \longversion{\ref{thm:g_off_all_TS_Pspace}}\shortversion{\ref{thm:all_All_TJ}}
       & \ref{cor:global-TS}&(\ref{lem:global-ind_off_tj_logspace})\\
       TJ & \shortversion{\ref{thm:all_All_TJ}}\longversion{\ref{thm:defall_TJ_PSpace}}
       & \shortversion{\ref{thm:all_All_TJ}}\longversion{\ref{thm:offall_TJ_PSpace}}
       &\longversion{\ref{thm:g_pow_all_TJ_Pspace}}\shortversion{\ref{thm:all_All_TJ}}
       &\shortversion{\ref{thm:all_All_TJ}}\longversion{\ref{thm:i_off_all_TJ_Pspace}}
       &\shortversion{\ref{thm:all_All_TJ}}\longversion{\ref{cor:globaldefall_TJ_PSpace}}
       & \longversion{\ref{thm:g_off_all_TS_Pspace}}\shortversion{\ref{thm:all_All_TJ}}
       & \longversion{\ref{thm:g_off_all_TS_Pspace}}\shortversion{\ref{thm:all_All_TJ}}&(\ref{lem:global-ind_off_tj_logspace})\\
       TAR & \ref{cor:all-reconf-TAR} & \ref{cor:all-reconf-TAR} & \ref{cor:all-reconf-TAR}
       & \ref{cor:all-reconf-TAR}
       & \ref{cor:all-reconf-TAR}
       & \ref{cor:all-reconf-TAR}
       & \ref{cor:all-reconf-TAR}
       & (\ref{cor:g_i_offensive_tar})
\end{tabular}
\caption{Survey on \pspace-completeness results for alliance reconfiguration problems (or membership in \logspace\ when put in parentheses).  \longversion{The references refer to results of our paper.}\label{tab:Alliance-Reconfig-complexity}}
\end{table}

We survey our classical complexity results in \autoref{tab:Alliance-Reconfig-complexity}. Notice that we alternate between \logspace- and \pspace-results.
Admittedly, our \FPT-algorithms are not optimized in terms of running times. As most of our arguments are of a combinatorial nature, one could also interpret these results as kernel results. Alternatively, one could construct branching algorithms that make use of our combinatorial findings. 
The parameterized complexity status of $(k+\ell)$-\offallianceReconf-TJ, $(k+\ell)$-\textsc{Idp}-\offallianceReconf-TJ, \nd-$X$-\longversion{\AllReconf}\shortversion{\textsc{Reconf}}, $k$-$X'$-\longversion{\AllReconf}\shortversion{\textsc{Reconf}}, where $X$ is any alliance condition and $X'=\{\textsc{\,\longversion{Def, Off, Idp-Off}\shortversion{DA,OA,Idp-OA}\,}\}$ is still open.

\begin{toappendix}
Considering the underlying combinatorial question of finding an alliance of cardinality at most~$k$, this is known to be \NP-complete for any of the discussed variants; see \cite{Cametal2006,Caretal2007,FerRai07,JamHedMcC2009,Sha2004}. For simple \FPT-results with parameterization by solution size, we refer to \cite{FerRai07}, while discussions on kernel sizes can be found in~\cite{FerFHKMN2020}. 
\end{toappendix}


\bibliographystyle{splncs04}
\longversion{\bibliography{alliance}}
\shortversion{\bibliography{alliance}}

\end{document}